\documentclass[preprint]{aastex}
\newcommand{\ba}{\beta_{app}}

\def\gtrsim{\mathrel{\hbox{\rlap{\hbox{\lower4pt\hbox{$\sim$}}}\hbox{$>$}}}}
\def\lesssim{\mathrel{\hbox{\rlap{\hbox{\lower4pt\hbox{$\sim$}}}\hbox{$<$}}}}

\shortauthors{Hogan et al.}
\shorttitle{Chandra Discovery of 10 New X-Ray Jets Associated With FR II Radio Core-Selected AGNs in the MOJAVE Sample}
\begin{document}
\slugcomment{}
\title{Chandra Discovery of 10 New X-Ray Jets Associated With FR II Radio Core-Selected AGNs in the MOJAVE Sample}
\author{Brandon Hogan$^1$, Matthew Lister$^1$, Preeti Kharb$^2$,\\
 Herman Marshall$^3$, Nathan Cooper$^1$}

\affil{$^1$Purdue University,\\
West Lafayette, IN 47907}
\affil{$^2$Rochester Institute of Technology,\\
Rochester, NY 14623}
\affil{$^3$MIT Kavli Institute,\\
Cambridge, MA 02139}

\email{hoganb@purdue.edu\\
%mlister@physics.purdue.edu\\
%pxksps@cis.rit.edu\\
%hermanm@space.mit.edu\\
%nukwaste@physics.purdue.edu
}

\begin{abstract}
The \textit{Chandra X-ray observatory} has proven to be a vital tool for studying high-energy emission processes in jets associated with Active Galactic Nuclei (AGN). We have compiled a sample of 27 AGN selected from the radio flux-limited MOJAVE (Monitoring of Jets in AGN with VLBA Experiments) sample of highly relativistically beamed jets to look for correlations between X-ray and radio emission on kiloparsec scales. The sample consists of all MOJAVE quasars which have over 100 mJy of extended radio emission at 1.4 GHz and a radio structure of at least 3$\arcsec$ in size. Previous \textit{Chandra} observations have revealed X-ray jets in 11 of 14 members of the sample, and we have carried out new observations of the remaining 13 sources. Of the latter, 10 have X-ray jets, bringing the overall detection rate to $\sim$ 78$\%$.  Our selection criteria, which is based on highly compact, relativistically beamed jet emission and large extended radio flux, thus provides an effective method of discovering new X-ray jets associated with AGN. The detected X-ray jet morphologies are generally well correlated with the radio emission, except for those displaying sharp bends in the radio band.  The X-ray emission mechanism for these powerful FR II (Fanaroff-Riley type II) jets can be interpreted as inverse Compton scattering off of cosmic microwave background (IC/CMB) photons by the electrons in the relativistic jets.  We derive viewing angles for the jets, assuming a non-bending, non-decelerating model, by using superluminal parsec scale speeds along with parameters derived from the inverse Compton X-ray model.  We use these angles to calculate best fit Doppler and bulk Lorentz factors for the jets, as well as their possible ranges, which leads to extreme values for the bulk Lorentz factor in some cases.  When both the non-bending and non-decelerating assumptions are relaxed the only constraints on the kpc scale jet from the Chandra and VLA observations are an upper limit on the viewing angle, and a lower limit on the bulk Lorentz factor.
\end{abstract}
%Thus, we also explore the ramifications of relaxing the non-bending and non-decelerating assumptions of the model.


\section{Introduction}
Blazar jets are generated in active galactic nuclei (AGN) as a result of accretion onto supermassive black holes, and can transport energy over large distances.  These outflows tend to show apparent superluminal speeds, and are oriented at very shallow angles with respect to the line of sight \citep{AS80}.  The blazar class encompasses flat spectrum radio quasars (FSRQs), which have Fanaroff-Riley type II jets (FR II; \citealt{FR74}), and BL Lac objects, which are thought to have FR I type jets \citep{UP95}.  The AGN outflows that we discuss here are of the powerful FR type II class which have well collimated jets and bright terminal hotspots (e.g., \citealt{PK08}).  In terms of X-ray production in the jet, the inverse Compton radiation process is suggested to be more important in FSRQs than in the less powerful BL Lac sources \citep{GT08,HK06}. 

Prior to the launch of the \textit{Chandra X-ray observatory}, there were very few AGN jet detections in the X-ray band.  The only major X-ray imaging telescopes in use were \textit{Einstein} and \textit{ROSAT}.  Only a few very bright, nearby X-ray jets were known, e.g., M87, 3C 273, Centaurus A, and a few lesser known sources (see \cite{RS04}, \cite{HM05} and references therein).  Since the launch of \textit{Chandra} there have been approximately 50 new discoveries of  X-ray jets that are spatially correlated to some extent with the radio emission.  The excellent angular resolution of \textit{Chandra} has revealed detailed structure in FR II jets, such as knots, lobes and hotspots \citep{HC02}, and has opened up an entirely new subfield of AGN astronomy.  

Many X-ray emitting jets were discovered in early surveys by \cite{RS04} \& \cite{HM05}.  The quasars in these surveys were selected mostly from radio imaging surveys of FSRQs, but the surveys were not statistically complete.  For our study we have chosen to assemble a complete sample of beamed FR II jets according to well defined selection criteria.  These jets generally have high Doppler factors and relativistic speeds.
The MOJAVE Chandra Sample (MCS) is a complete subset of compact radio jets selected from the MOJAVE sample.  The latter sample consists of all 135 known AGN with $\delta$ $>$ $-20^\circ$, $|$b$|$ $>$ 2.5$^\circ$ and VLBA 15 $\sim$ GHz correlated flux density exceeding 1.5 Jy at any epoch between 1994.0 and 2004.0 (2 Jy for AGN below $\delta = 0\arcdeg$) \citep{ML09a}.  Since the long interferometric baselines of the VLBA are insensitive to large-scale unbeamed radio emission, the sample is heavily dominated by blazars.  In Section 2, we describe how the MCS was selected using a set of criteria designed to maximize the chances of X-ray jet detection.

The goals of our study are threefold.  First, we seek to identify new X-ray jets for future follow up with \textit{Chandra}, \textit{Spitzer}, and the \textit{Hubble Space Telescope (HST)}.  Because of the large redshift range of the MCS (0.033 $\leq$ z $\leq$ 2.099), we can examine the effects of proposed X-ray mechanisms such as inverse Compton scattering off of cosmic microwave background (IC/CMB) photons by relativistic electrons in the jets, which is highly dependent on redshift.  Second, we wish to characterize the ratio of X-ray emission to radio emission for a large complete sample of jets.  Such information is vital for determining the respective roles that deceleration and bending play in determining why jets associated with some AGN are strong X-ray emitters.  Lastly, we can use the detailed viewing angle and speed information of the AGN jets on parsec (pc) scales provided by the MOJAVE program to better model the X-ray emission mechanism(s).  

A total of 14 AGN in the MCS have previously been observed by \textit{Chandra} \citep{HM05,RS04,CO02,SJ06,JM06,WE07,WA01}; here we present new 10 kilosecond exposures on the other 13 sources, in which we have detected 10 new X-ray jets.  A detailed analysis of the full 27 source sample will be presented in subsequent papers.  

Our paper is laid out as follows; we describe the MOJAVE Chandra Sample in Section 2, along with our data reduction method and selection criteria.  In Section 3, we describe the jet observations for each specific source in which a jet was present in both the radio and X-ray images.  In Section 4, we discuss overall source trends and provide additional ancillary information on selected sources.  In Section 5, we discuss implications of the model with respect to the bulk Lorentz factor and viewing angle.  We summarize our conclusions in Section 6.  The limits for the derived Doppler factor and bulk Lorentz factors are given in the Appendix.  Throughout this paper we use a standard cosmology with H$_0$ = 71 km s$^{-1}$ Mpc$^{-1}$, $\Omega$$_m$ = 0.27, and $\Omega$$_{\Lambda}$ = 0.73.

\section{The \itshape{MOJAVE Chandra} Sample}

\subsection{Selection Criteria}
In formulating the sample for our \textit{Chandra} survey, we wished to focus on relativistic radio galaxy and blazar jets whose high Doppler factors would make them prime candidates for IC/CMB X-ray emission.  We also decided to limit our survey to FR II radio galaxies and quasars, in order to avoid possible contamination by lower power (presumably FR-I type) BL Lac objects.  The MOJAVE sample \citep{ML09b} provided a useful list of candidates for possible IC/CMB in this regard, since it comprises a complete set of compact radio jets in the northern sky.  Its VLBA selection criteria favor highly Doppler-boosted blazar jets for which extensive pc scale kinematic information has been obtained \citep{ML09a}.  Deep 1.4 GHz VLA A-configuration radio images are also available for the entire sample of 135 AGN \citep{NC07,PK10}.  In order to maximize the likelihood of X-ray jet detection, we considered all MOJAVE quasars and FR II radio galaxies having more than 100 mJy of extended kiloparsec (kpc) scale emission (where the extended emission is the total emission after the core emission has been removed) at 1.4 GHz and a radio structure that was at least 3$\arcsec$ in extent.  This final list of 27 AGN (see Table \ref{table:mcs}) comprises the MCS.  A search of the \textit{Chandra} archive revealed that 14 of these objects had been previously observed, where most of the sources had integration times $>$ 10 kiloseconds.  During the period 2007, November to 2008, December, we obtained new \textit{Chandra} 10 kilosecond ACIS images of the remaining 13 AGN.

\subsection{Data Reduction and Analysis}

\begin {figure}[t]
\begin {center}
\includegraphics [scale=.5] {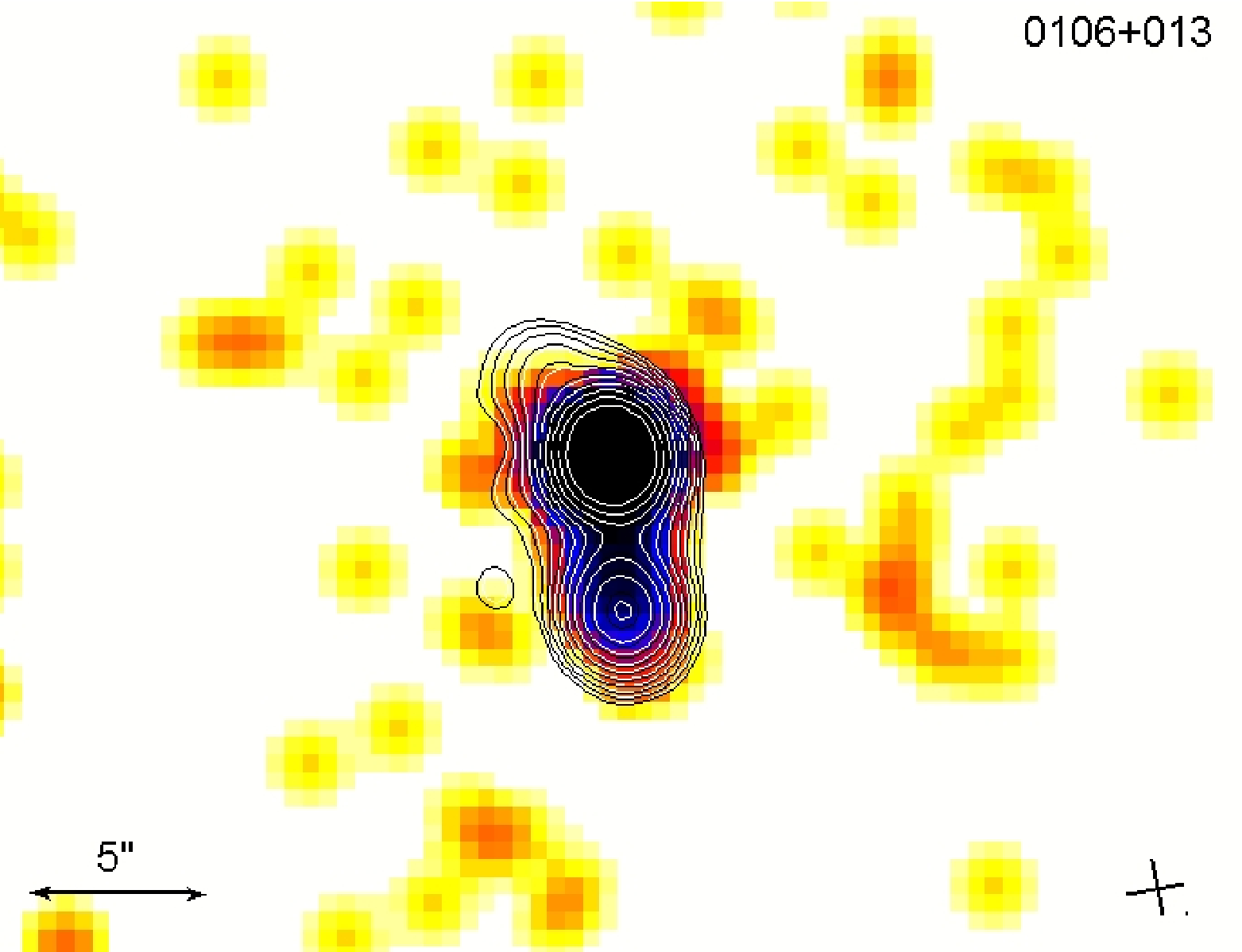}
\caption{X-ray images obtained from \textit{Chandra} with VLA 1.4 GHz radio contours overlaid in black and white.  The VLA contours are set at 5 times the rms noise level for the lowest contour, with the exception of 0415+379 and 1849+670, which had their starting values set to 10 and 2.5 times the rms noise respectively, and multiples of 2 greater than that for each successive level.  The X-ray portion of each image has been energy filtered to a range of 0.5 to 7.0 keV in CIAO before being processed in DS9.  The FWHM dimensions of the radio restoring beam are denoted by a cross in the bottom corner of each image and are also located in Table \ref{table:vla}.}
\label{fig:rxo}
\end{center}
\end{figure}

In Figure \ref{fig:rxo} we present X-ray-radio overlays for our new observations.  We first obtained the 1.4 GHz VLA A-array data from the NRAO\footnote{The National Radio Astronomy Observatory is a facility of the National Science Foundation operated under cooperative agreement by Associated Universities, Inc.} data archive and our own observations \citep{NC09,PK10}.  The observation dates and exposure times for the new \textit{Chandra} targets are listed in Table \ref{table:obs}.  These were reduced following standard procedures in the Astronomical Image Processing System (AIPS).  After the initial amplitude and phase calibration using the standard calibrators, the AIPS tasks CALIB and IMAGR were used iteratively to self-calibrate and image the sources.  Self-calibration on both the phases (with solution intervals typically set to less than 0.5 mins in CALIB) and amplitude (with successively decreasing solution intervals) were performed until convergence in image flux and structure was achieved.  The final radio maps had a typical rms noise of $\sim 0.2$~mJy~beam$^{-1}$.  The FWHM restoring beam of the radio images was adjusted to $\sim 1.4\arcsec$ on average, and the width of the \textit{Chandra} FWHM was estimated to be $\sim$ 0.75$\arcsec$.  The \textit{Chandra} maps for the X-ray-radio overlays were created using the DS9 image tool.  We started by loading the level 2 events files into CIAO for energy filtering.  The event files were filtered to an energy range of 0.5 to 7 keV.  After loading these event files into DS9, we used the analysis smoothing tool to smooth the image.  This was set to use Gaussian smoothing with a kernel radius of three pixels, where a pixel size of 0.5 pixels per arcsecond are used.  After that we adjusted the color scale so that the cores were oversaturated.  This allowed for the jet emission to be detected easily by visual inspection.  The radio contours were then superimposed.  The images frames were aligned in DS9 using the WCS frame matching setting.  Some of the core positions were slightly misaligned when overlayed.  We registered the X-ray and radio images using the Fv program in the Ftools package provided by NASA\footnote{Information about Ftools can be found at http://heasarc.gsfc.nasa.gov/ftools/}\citep{BB95}.  The shifts were generally small, of order of 2 pixels or less.\\  

The centroid positions were also calculated via the method described by \cite{HM05}.  The procedure involves determining the preliminary X-ray core centroid location by fitting Gaussians to the 1D histograms obtained from events within 30$\arcsec$ of the core region.  This defines the rough centroid position.  This process was then repeated using a region defined by a radius of 3$\arcsec$ from the previously calculated centroid position which allows for a more refined centroid position.  This two step approach reduces the effect of the extended jet emission, which can bias the centroid.  After calculating the centroid position, we used Poisson statistics to test for the existence of an X-ray jet.  The radio profiles were first used to define the outer radius and position angle of the primary jet from visual inspection, which are listed in Table \ref{table:vla}.  The radii allowed us to create a box within which we could check for the existence of X-ray structures.  The lengths of the radii varied but the width of each rectangle was fixed at 3$\arcsec$.  The inner radius was fixed at 1.5$\arcsec$ to eliminate the core emission, with the exception of 1849+670 and 2345$-$167, which were fixed to values of 5$\arcsec$ and 2$\arcsec$ respectively, because of their elongated radio restoring beams (Table \ref{table:vla}).

\begin{figure}
\includegraphics {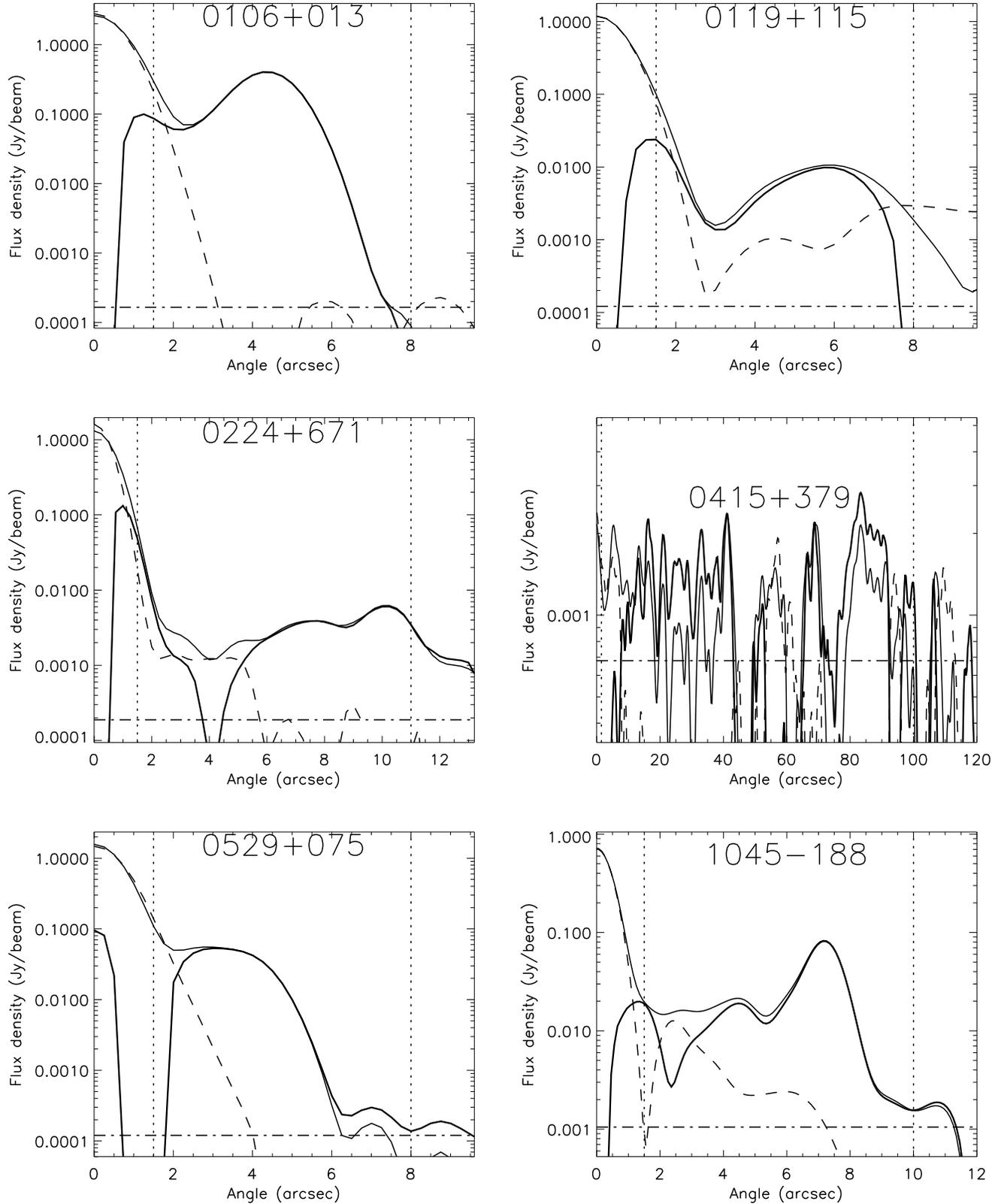}
\caption{\tiny{Radio profiles for sources in the sample.  The thin solid lines give the radio profiles along the position angle of the jets.  This information is given in Table \ref{table:sjm} and is used in the measurements of the X-ray jet emission.  The dashed lines indicate the radio profile at a position angle of 90$^\circ$ counter-clockwise from the jet to avoid any non-jet emission and counter jet emission.  The solid, bold line indicates the difference between the two profiles so that core emission is removed and the effective flux can be measured.  The horizontal dot-dashed lines are set to a value five times the average noise level and the vertical dashed lines show the inner and outer radius limits.}}
\label{fig:rp}
\end{figure}

\begin{figure}
\includegraphics {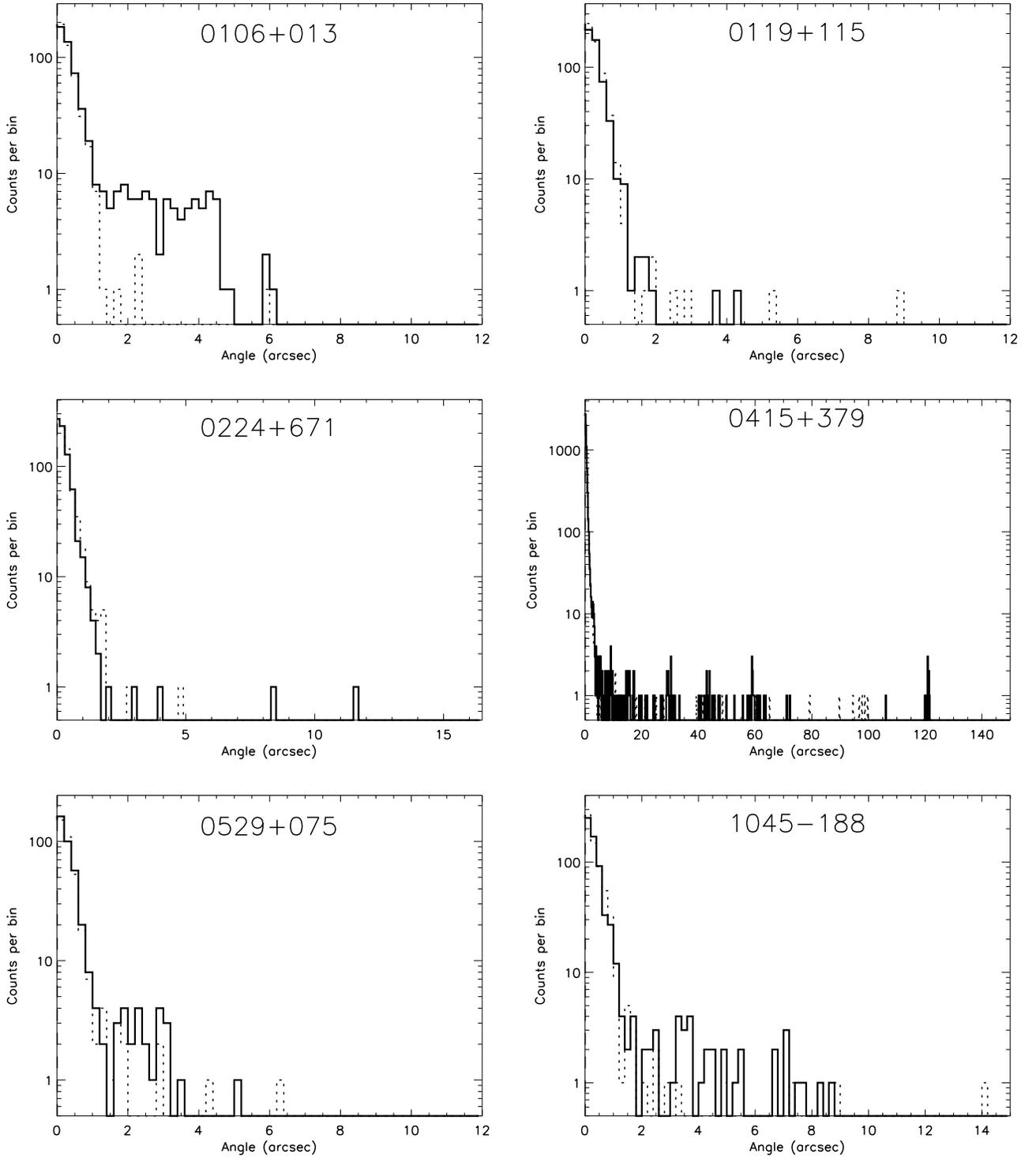}
\caption{X-ray profiles for the sources in the sample.  These are represented as histograms of the counts in 0.2$\arcsec$ bins.  The solid lines give the profile along the position angle of the jet, as defined by the radio images.  The dashed lines show the profile along the counter-jet direction, which is defined as 180$^\circ$ opposite to the jet.}
\label{fig:xp}
\end{figure}

The detection algorithm assumes straight radio jets and that the region 90$^\circ$ to the primary jet is free of jet emission, which is a valid assumption for all of the jets in the sample except for 0119+115. This source has radio jet emission on both sides of the core perpendicular to the direction of the primary jet but lacks substantial X-ray emission, making the perpendicular radio jet emission irrelevant.  We then produced profiles of the radio emission along the jet position angle, and at 90$^{\circ}$ to it (Figure \ref{fig:rp}).  These two quantities were subtracted from each other to eliminate core structure.  The X-ray jet counts were then compared for the same sky region.  For the X-ray profiles, we chose to use the jet axis region and the region 180$^\circ$ away from it (Figure \ref{fig:xp}).  We then compared the counts in these regions by using Poisson statistics.  Sources with negative values in the count rate column in Table \ref{table:sjm} have less X-ray emission compared to the area in the region opposite to it.  Counter-jets in powerful AGN have rarely been seen in X-rays, presumably due to Doppler boosting effects \citep{WD09} and we found no evidence of any X-ray counterjets in our sample from visual inspection.  We set the Poisson probability threshold for the detection of an X-ray jet to 0.0025 \citep{HM05}.  This value yields a 5\% chance of a false detection in one out of every 20 sources.  The X-ray fluxes were computed from count rates using a conversion factor of 1 $\mu$Jy per count s$^{-1}$.  This conversion is accurate to about 10\% for typical jet power law spectra \citep{HM05}.  Our analysis method indicated that there were X-ray jet detections in all of the sample sources except for 0119+115, 0224+671, and 2345$-$167.  These sources did not show any appreciable X-ray emission above the background level except for their respective cores (Figure \ref{fig:rxo}).  This is despite the fact that their redshift and radio structure are comparable to the other sources in the sample.  The relevant X-ray emission limits for these sources are listed in Tables \ref{table:sjm} and \ref{table:bmp}.

\section{Notes on Individual Sources}
In this section we provide a general overview of the X-ray jet morphologies in our sample, and how they compare with the 1.4 GHz radio structure seen in the VLA A array radio images (Figure \ref{fig:rxo}).  We use the term ``knot'' for any excess of emission at a shock front that is not at the terminal point of the jet, and use the term ``hotspot'' for any knot-like structure or excess of emission located where the jet terminates in the radio band.  Some of these images may show readout streaks which look similar to jet emission.  We attempted to set the roll angle for each source that we viewed with \textit{Chandra} so that the readout streak would not be aligned with the previously known jet emission in the radio band. We have labeled these in Figure \ref{fig:rxo}.  Figures \ref{fig:rp} and \ref{fig:xp} show the radio and X-ray jet profiles, respectively, and are described in section 2.2.  

\subsection{0106$+$013 (OC 12)}
This blazar has a prominent X-ray jet which protrudes due south of the radio core.  There is a strong correlation between the X-ray jet and the radio jet contours.  Both have a hotspot-like structure 5$\arcsec$ from the core, which is the terminal point of their respective jets.  There is also an excess of radio emission to the northeast that does not correlate with any X-ray emission.  
 
\subsection{0415$+$379 (3C 111)}
The VLA radio data on this powerful radio galaxy was obtained by \cite{LP84}.  The image of the jet shows that there are 4 prominent radio knots present, and three of these show an excess of X-ray emission.  The terminal knot, or hotspot, shows an excess of emission also, indicating an excellent correlation between the X-ray and radio emission in this jet.  The jet lies at a reasonably small angle to the line of sight according to our IC/CMB calculations (see Section 4 and Table \ref{table:bmp}).  \cite{SJ05} give a value of the angle to the line of sight of 18.1 $\pm$ 5.0$^\circ$, which is somewhat larger than our value obtained via the IC/CMB method.  This jet has a measured superluminal speed of 5.9c and therefore has a maximum value for the angle to the line of sight of roughly 19$^\circ$ \citep{ML09a}.  The deprojected length from the core to the terminal hotspot for this source is 661 kpc for $\theta = 8^\circ$, 537 kpc for $\theta = 10^\circ$, and 302 kpc for $\theta = 18^\circ$, where the measured distance on the plane of the sky is $\sim$100$\arcsec$ from the core to the hotspot.

This object also shows only a single sided jet structure within 100$\arcsec$ of the core, which is likely to be the result of Doppler boosting and also shows a pc scale jet in approximately the same general direction as the kpc jet in the radio band.  The terminal point of both the jet and counter jet are visible, with lobe-like radio emission at the respective terminal points, along with notable X-ray emission only at the primary jet terminal point.  This object is one of only two radio galaxies, the other being Cygnus A, that met the selection criteria of the MCS and have X-ray jet structure \citep{WA01}.

\subsection{0529$+$075 (OG 050)}
This blazar has an X-ray jet which follows the initial radio jet position angle of $-$145$^\circ$.  The radio jet terminates at $\sim$ 6$\arcsec$ 
from the core.  There is also radio emission to the southeast of the core and X-ray emission more to the east of the core, but these two features are not coincident.  The main jet emission seems to coincide with the X-ray jet in the south west direction.  The south eastern emission could be coming from the counter-jet, since we see no X-ray emission there, which is what we might expect from the Doppler de-boosting of a counter jet, according to the IC/CMB model.  The pc scale jet lies at a position angle of approximately $-$45$^\circ$, which implies significant bending between the pc and kpc scale jets.

\subsection{1045$-$188}
The radio image of this blazar shows strong one sided jet emission and diffuse lobe emission from the counter jet.  The main jet, which lies at a position angle of 125$^\circ$, makes a $\sim$ 90$^\circ$ bend at the bright radio knot at a distance of $\sim$ 8$\arcsec$ from the core.  The X-ray emission follows the primary radio jet out to the bend but then terminates abruptly.  There is no detectable X-ray emission from the counter jet or its associated lobe above the background level. 
 
\subsection{1334$-$127}
This blazar has an X-ray jet with a length of $\sim$ 6$\arcsec$ that follows the radio jet emission out to a 60$^{\circ}$ bend of the radio jet, then undergoes a drop in emission, but still terminates at the same point as the radio jet.  Both jets initially follow a position angle of 135$^\circ$.  The emission characteristics in the bend region are significantly different than the jet of 1045$-$188, which undergoes a sudden drop in X-ray emission after the bend.

\subsection{1800$+$440}
The radio image of this jet shows a lobe to the northeast, and emission toward the southwest at a position angle of $-$130$^\circ$.  The X-ray jet emission follows the radio jet emission to the southwest for $\sim$ 4$\arcsec$ and then terminates at the radio knot.  The radio emission continues for another $\sim$ 3$\arcsec$ until it terminates in a hotspot.  There also appears to be a shallow bend beyond the first radio knot.  This is another example where the X-ray jet flux decreases beyond a radio knot located at a bend.  

\subsection{1849+670}
This quasar shows a short radio jet to the north with lobe structure to the south.  The radio emission to the north shows a drop in emission at $\sim$ 15$\arcsec$, whereas the X-ray emission decreases drastically after a distance of about 9$\arcsec$ from the core.  A visual inspection of the overlays (Figure \ref{fig:rxo}), indicates no apparent correlation between the radio counter lobe emission and the X-ray emission in this source.   

\subsection{2155$-$152}
Although there is arcsecond scale radio emission on both sides of the core, the pc scale jet is oriented to the south, so we chose this direction to search for X-ray emission.  The southern jet emission terminates at a distance of $\sim$ 8$\arcsec$ from the core, whereas the X-ray emission is seen up until a distance of $\sim$ 4$\arcsec$ south of the core where it terminates.  

\subsection{2201$+$315 (4C 31.63)}
The radio map of this object shows a $\sim$ 37$\arcsec$ long jet with a counter jet lobe located approximately the same distance away in the opposite direction.  There is X-ray emission which correlates with the radio in the first knot structure as seen in Figures \ref{fig:rp} and \ref{fig:xp}.  This is at a distance of $\sim$ 4$\arcsec$ from the core.  Downstream from this knot, there is a significant decrease in X-ray flux.  This detection is considered marginal and needs a longer exposure time to be able to produce a clear visual correlation in the overlay images.     

\subsection{2216$-$038}
The radio image shows a sharp knot-like structure at a distance of about 8$\arcsec$ from the core, along with an excess in the X-ray emission at the same location.  The radio jet shows emission continuing out to a distance of $\sim$ 15$\arcsec$, but the overlayed image shows a sharp decrease in X-ray emission between the knot and the terminal hotspot.  There is a counter jet lobe to the northeast in the radio portion of the overlay, but there is no X-ray emission in this region above the background emission.

\section{X-ray Data Analysis}

\subsection{Data Trends and Results}

We have found that the extended flux densities, S$_{ext}$, are closely correlated with the detection rate of the X-ray emission.  Interestingly, \cite{PK10} have reported a significant trend between pc scale (apparent) jet speeds and extended radio luminosity in the MOJAVE blazars.  This could suggest a link between X-ray jet detection and jet speed.  Although our selection criteria guaranteed S$_{ext}$ $>$ 100 mJy for all sources, we get a 100\% X-ray jet detection fraction for S$_{ext}$ $>$ 300 mJy.  Below that value, we find a significantly lower detection rate ($\sim$ 57\%).  Using an extended flux density threshold value as a selection criterion could prove to be a definitive way to predict X-ray jet detection in blazars and bright core radio galaxies from radio jet images alone.\\

We also ran Kolmogorov-Smirnov tests on the MCS population for three different cases; the $\beta_{app}$ values with respect to the detection of sources, the $\beta_{app}$ values with respect the S$_{ext}$ threshold value (300 mJy), and the redshift value with respect to the detection of the sources.  In all three cases, the \textit{p} value does not reject the possibility that both populations could have the same parent population.
 
If we examine the X-ray jet morphologies of the MCS, as well as the sample of \cite{HM05}, we find no visible instances of any X-ray counterjets.  This is consistent with the predictions of the IC/CMB model, which suggest that the X-ray emission from the relativistic kpc scale jets must be highly beamed along the direction of the flow.  Thus, any X-ray counterjet emission would not be visible to \textit{Chandra}.  Jet bending on the kpc scale can also limit the detection of X-ray jets that are bright in the radio band.  This could occur if there is deceleration past the point of the bend, which may in turn limit the flow to mildly relativistic speeds.  A secondary factor may be a significant change in the magnetic field strength downstream from the bend as in 3C 279 \citep{SJ04}.  Other examples of jet bending where the X-ray flux decreases past the bend in the MCS are the blazars 0234$+$285, 1045$-$188, 1222$+$216, 1334$-$127, 1510$-$089, 1800$+$440, and 1928$+$738.

\subsection{ Emission Modeling}
The MCS has improved the detection rate of X-ray jets when compared to previous surveys of the same kind.  We have found that $\sim$ 78\% (21 of 27) of the objects in the MCS have significant kpc scale X-ray jet emission in 10 kilosecond \textit{Chandra} exposures, while the previous surveys have a $\sim$ 60\% detection rate \citep{HM05,RS04}.  One difference between previous quasar surveys and ours is that the MCS is complete with respect to beamed synchrotron emission.  Our sample is also more strongly biased toward blazars with superluminal jet speeds and higher Doppler factors.  Our detection rate supports the hypothesis that relativistic beaming is indeed an important factor affecting large-scale X-ray jet emission.\\

For our sample analysis of blazars and radio galaxies, we have chosen to use a method similar to the one used by \cite{HM05}.  This model assumes that the magnetic fields in the kpc scale jets are in equipartition with the particle energies.  Under this assumption, the magnetic field strength of the jet can be calculated.  The photon energy density is then compared to the magnetic field energy density by using the previously calculated magnetic field strength and the CMB photon density.  The parameter \textit{K} (a function of $\beta$, the jet velocity in terms of the speed of light, and $\theta$, the angle to the line of sight), can be calculated by combining the magnetic field strength with the ratio of observed X-ray (inverse Compton) to radio (synchrotron) luminosities.

Since the single component synchrotron model has difficulties in explaining the X-ray emission in powerful blazar jets that have observed limits on their optical emission (e.g., \citealt{HM05}), we have derived physical quantities for the X-ray emission using a standard IC/CMB model.  We started with the same IC/CMB assumptions as \cite{HM05}, which were obtained from \cite{HK02}.  The first assumption is that the energy density of the CMB occurs at the peak of the blackbody distribution.  The second assumption is that the jet frame equipartition holds between the particle energy densities and the magnetic field, with a filling factor ($\phi$) of 1.  If relativistic protons contribute to the particle energy density, then this assumption will fail and beaming will become much more extreme.  The third assumption is that the low energy spectral index for the synchrotron spectrum continues unchanged below the current range of the instruments used to measure them.  The procedure involves first defining

\begin{equation}
    B_{1} = \left[\frac{18.85C_{12}(1+k)L_{s}}{\Phi V}\right]^{2/7},
\end{equation}

\noindent where \textit{$B_{1}$} is the spatially averaged, minimum energy magnetic field of the jet in the case where there is no Doppler boosting, in Gauss, \textit{$C_{12}$} is a weak function of the low frequency spectral index of the synchrotron spectrum ($\alpha_r$, where S$_\nu$ $\propto$ $\nu$$^{-\alpha_r}$), $\Phi$ is the filling factor, $L_{s}$ is the synchrotron luminosity (calculated from the radio flux and luminosity distance), \textit{k} is the baryon energy fraction parameter, and \textit{V} is the emitting volume \citep{AP70}.  The values used for the constants are; \textit{k}=0, \textit{$C_{12}$}=5.7$\times$10$^7$, $\alpha_r$=0.8, and $\Phi$=1.  The emitting volume is calculated using the length of the jet defined in Table \ref{table:sjm} by taking the difference of the two radius values and then assuming a cylindrical cross section given by the width associated with the \textit{Chandra} FWHM (0.75$\arcsec$).  The 1.4 GHz (FWHM = 1.4$\arcsec$) radio data results in larger derived emitting volumes than the \textit{Chandra} FWHM.  This discrepancy causes the magnetic field value (\textit{$B_{1}$}) to be considered a minimum value for all intents and purposes.  One way to address this magnetic field discrepancy is to adjust the filling factor $\Phi$.  If the $\Phi$ is decreased from unity by a factor of 10 the magnetic field quantity $B_{1}$ would only increase by roughly a factor of 2 \citep{HM05}.

We next compute the X-ray to radio luminosity ratio (R) using

\begin{equation}
R = \frac{S_{x}(\nu/\nu_{x})^{-\alpha_r}}{S_{r}(\nu/\nu_{r})^{-\alpha_r}} = \frac{S_{x}\nu^{\alpha_r}_{x}}{S_{r}\nu^{\alpha_r}_{r}} = \left[\frac{\nu_{x}}{\nu_{r}}\right]^{\alpha_r-\alpha_{rx}},
\end{equation}

\noindent where $\nu_{r}$ and $\nu_{x}$ are the radio and X-ray frequencies at which the flux densities $S_{r}$ and $S_{x}$ are observed, respectively.  The luminosity distance and redshift are also important parameters, because they affect the derived synchrotron luminosity in the jet frame.  Equation 2 is valid as long as both the X-ray and radio frequencies are far from the endpoints of the synchrotron and IC spectral breaks.  We use $\nu$$_x$=2.42$\times$10$^{17}$ Hz for calculation purposes later in the paper, as well as the $\nu$$_r$ values listed in Table \ref{table:sjm}.  Based on previous IC/CMB modeling, and following the solution presented by \cite{HM05}, we compute the quantity \textit{K}, which is a function of constants or observed values only, given by

\begin{equation}
K=B_{1}(aR)^{1/(\alpha_r+1)}(1+z)^{-(\alpha_r+3)/(\alpha_r+1)}b^{(1-\alpha_r)/(\alpha_r+1)},
\end{equation}

\noindent where \textit{a}=9.947$\times$10$^{10}$ Gauss$^{-2}$ and \textit{b}=3.808$\times$10$^{4}$ Gauss, as used by \cite{HK02}.  The values for \textit{a} and \textit{b} are found by equating the expected and observed values of the ratio of X-ray to radio energy densities (\textit{R}) under the equipartition assumption.  This leads to \textit{K} being a dimensionless number that is solely a function of intrinsic jet speed and viewing angle.  \cite{HM05} showed that \textit{K} is a simple function of the beaming parameters under the assumption that $\Gamma > 1.5$.

\begin{equation}
K=\frac{1-\beta+\mu-\beta\mu}{(1-\beta\mu)^{2}},
\label{eq:K_eqn}
\end{equation}

\noindent which can then be solved for $\mu$ (where $\mu$ = cos $\theta$) for a given $\beta$ (see \cite{HM05} Eq. 5).  The solution for $\mu$ has two roots and we have chosen, like \cite{HM05}, to use the negative root of Equation \ref{eq:K_eqn2}.
\begin{equation}
\mu=\frac{1-\beta+2K\beta\pm(1-2\beta+4K\beta+\beta^{2}-4K\beta^{3})^{1/2}}{2K\beta^{2}}
\label{eq:K_eqn2}
\end{equation}

\noindent If the viewing angle is known, the following equations can be used to find $\delta$ and $\Gamma$:

\begin{equation}
\beta=\frac{\ba}{\ba\mu+\sqrt{1-\mu^2}},
\label{eq:beta_app_eqn}
\end{equation}

\begin{equation}
\theta=\arctan\frac{2\beta_{app}}{\beta^2_{app}+\delta^2-1},
\end{equation}

\begin{equation}
\Gamma=\frac{\beta^2_{app}+\delta^2+1}{2\delta}.
\end{equation}

By examining the effect of each parameter on the K factor individually  we have determined that the main source of observational error in \textit{K} is the radio spectral index $\alpha_{r}$.  Typical observed values for $\alpha_{r}$ in kpc scale jets are between $-$0.7 and $-$0.9.  Since we do not have direct measurements of $\alpha_{r}$ for our jet sample, we carried out a Monte Carlo error analysis, using a Gaussian distribution of $\alpha_{r}$ with $\overline{\alpha_{r}}$ = $-$0.8 and $\sigma_{\alpha_{r}}$ = 0.1.  We tabulate the resulting one sigma error values for \textit{K} in Table \ref{table:bmp}.

\section{Viewing Angle and Bulk Lorentz Factor Analysis}

Using VLBI observations, we can investigate the possible kpc-scale jet beaming parameters under an initial assumption that there is no deceleration or bending from pc to kpc scales.  This was done by using the theoretical framework of \cite{HK02}.  Given a pc-scale $\ba$ measurement, which is a function of $\theta$ and $\beta$, we can use the \textit{K} and $\ba$ equations, along with the assumption that the value of $\ba$ is the same for the pc and kpc scale jets to solve for the viewing angle $\theta$.  This then allows us to calculate the Doppler factor ($\delta$) and the bulk Lorentz factor ($\Gamma$) using the IC/CMB model (e.g., see \citealt{HK02}).  We discuss the possibilities of jet bending and deceleration on the pc to kpc scales in Sections 5.2 and 5.3.

\subsection{IC/CMB model with No Jet Deceleration or Bending}
\begin{figure}
\includegraphics [scale=.4]{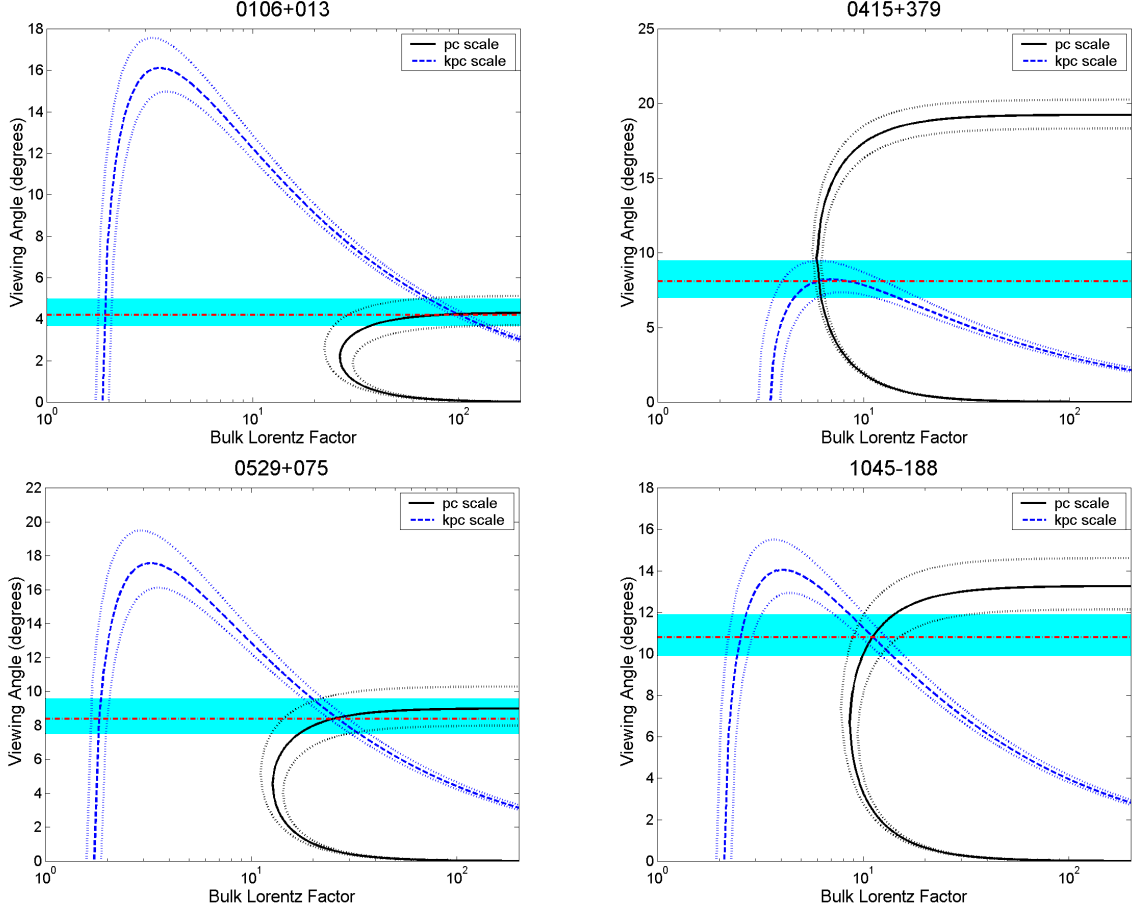}
\caption{\small{Plot of viewing angle ($\theta$) versus bulk Lorentz factor ($\Gamma$) for the jet.  The black curve (pc scale) is defined by Equation \ref{eq:beta_app_eqn}, using the observed value of the pc scale $\ba$ \citep{ML09b}.  The blue curve (kpc scale) is defined by Equation \ref{eq:K_eqn}, using the observed value for K (Table \ref{table:bmp}). The intersection point of the curves represents the jet Lorentz factor and viewing angle under the assumption of no bending or deceleration from pc to kpc scales (red dashed line).  Allowing for the possibility of deceleration, the kpc scale bulk Lorentz factor is given by the intersection of the red dashed line with the tail of the blue K curve at low $\Gamma$.  The cyan shaded region represents the possible range of $\theta$ if the jet decelerates from pc to kpc scales but does not bend. }}
\label{fig:gt1}
\end{figure}

Equation 4 can be solved for $\theta$ as a function of \textit{K} and $\beta$ (or $\Gamma$; \citealt{HM05}) as shown in Figure \ref{fig:gt1} (blue curve) for each source.  Equation 5 defines a locus of allowed $\Gamma$ and $\theta$ values for a fixed $\ba$ observed in the pc scale jet (black curve).  The intersection point of these curves yields the viewing angle and bulk Lorentz factor of the kpc scale jet, under the assumption that the X-ray emission is given by the IC/CMB model and that the jet directions and bulk Lorentz factors on pc and kpc scales are the same.  The range of the error of the $\ba$ and \textit{K} values defines a range for the value of $\Gamma$ described by the error curves associated with curves plotted in Figure \ref{fig:gt1}.  Note that in the cases of 0415+379 and 1800+440, as well as some other jets in the sample, the uncertainty in $\ba$ can translate into a large range of uncertainty on $\Gamma$ (Table \ref{table:bmp}).   

Our measured ranges of $\Gamma$ are consistent with previous investigations of beamed inverse Compton models for X-ray emission, which often require bulk Lorentz factors on the order of $\Gamma$$\approx$10 or greater.  \cite{LM09}, on the other hand, model a set of radio data using a Bayesian parameter-inference method, which provides $\Gamma$ values for a sample of FRII jets.  These $\Gamma$ values range from 1.18 to 1.49, which are significantly lower than the values required by the inverse Compton model.  The FRII jets in the \cite{LM09} sample, however, are selected on the basis of isotropic lobe emission, and thus their jets generally have large angles to the line of sight.  They are therefore more representative of the general FRII population than our MCS sample, which is highly biased toward fast jets pointing nearly directly at us.  As pointed out by \cite{LM97}, unbiased orientation samples of radio jets are likely to have much lower bulk Lorentz factors than blazars, due to the relatively steep power law distribution of jet speeds in the parent population.  \cite{NC10} produced a Monte Carlo simulation which describes the mean pc scale viewing angle distribution for a modeled MOJAVE sample.  The sample is modeled by using 1000 trial populations of 135 sources which have their bulk Lorentz factor described by a power law ranging from 3 to 50 with an index of $-$1.5 and are based on the luminosity function for the MOJAVE parent population \citep{CL08}. This simulation produces a roughly Poisson distribution of pc scale angles for the sample which is peaked around 2$^{\circ}$.  This non-uniform distribution for the pc scale viewing angle is produced because of the highly beamed nature of the MOJAVE sample.  Since the MCS is a subsample of the MOJAVE sample we should expect to see a similar angle bias in it.

0106+013 and 1849+670 show extreme values for the bulk Lorentz factor (71 $<$ $\Gamma$ $<$ 133 and 97 $<$ $\Gamma$ $<$ 129 respectively) when compared to the rest of the MCS sample, as well as other samples of blazars.  The largest value of $\Gamma$ on pc scales in the \cite{TH09} sample is 65 for 1730$-$130, which has an extremely large apparent velocity value ($\beta_{app}$ $\approx$ 35c).  \cite{TH09} compare their sample to the \cite{PU92} sample, which has a maximum $\Gamma$ of about 40.  The MOJAVE sample contains no known jets with superluminal speeds above 50c \citep{ML09b}.  \cite{LM97} find that in large flux limited blazar samples the value for $\beta_{app,max}$ should always be very similar to the $\Gamma_{max}$ in the parent population.  The MCS sources with extreme Lorentz factors have the smallest values of $\theta$ in the sample, with values less than 7$^{\circ}$, and also have the largest $\beta_{app}$ values in the sample.  

Based on papers by \cite{CM93} \& \cite{PM81} the small angle approximation with respect to the pc scale position angle ($\theta_{n}$) and the intrinsic misalignment angle between the pc and kpc scale jets ($\zeta$) for Equation 1 of \cite{CM93} is:

\begin{equation}
\tan(\eta)\approx\frac{\sin(\phi)}{\left(\frac{\theta_{n}}{\zeta}+\cos(\phi)\right)},
\end{equation}

\noindent where $\eta$ is the change in the position angle on the plane of the sky between the pc scale to the kpc scale jets and $\phi$ is the azimuthal angle of the jet.  For a small value of $\eta$, the value for $\zeta$ has to be small when compared to the value for $\theta_{n}$ for an arbitrary value of $\phi$.  Thus, for the sources in our sample with a small $\eta$, misalignments are likely to be less than a degree, so any discrepancy between $\delta$ and $\Gamma$ is likely to require deceleration.  Furthermore, to obtain a value for $\eta$ which is large, $\zeta$ must be larger than $\theta_{n}$.  These large values for $\eta$ require that the value for $\zeta$ is comparable to or larger than $\ba^{-1}$.  This is easily accomplished when the value for $\ba$ is large, as is true with most of the sources in this sample.  \cite{PM81} states that it is quite likely for small values of $\theta_{max}$ to be attributed to large values of $\eta$, where $\theta_{max}$ is the largest value of $\theta_{n}$ which is likely to occur.  Lastly, for sources with $\eta$ values which approach $90^{\circ}$ the value of $\zeta$ must be comparable to the value of $\theta_{n}$.  Examining Figure \ref{fig:gt1}, it is evident that bending between pc and kpc scales cannot by itself resolve the high bulk Lorentz factor issue in these two extreme sources.

\subsection{IC/CMB model with Jet Deceleration}
One way to reconcile the large $\Gamma$ values is to consider possible deceleration from the pc to kpc scale, where the deceleration is caused by a jet depositing its power into the surrounding medium in the form of kinetic energy \citep{GK04}.  A one-zone model deceleration of jets can allow for a misalignment of knots and other jet structures between the radio and X-ray wavelengths, which we also find examples of in the MCS.  Deceleration can also offer a way to reduce the unusually large values for $\Gamma$ in sources in the MCS by widening the beaming cone, assuming that the jets decelerate from ultra relativistic speeds near the base of the jet to mildly relativistic and even sub-relativistic speeds at the point of termination.  

We now examine the application of the IC/CMB model allowing for possible deceleration, but no bending between the pc and kpc scales.  The problem reduces to finding a possible family of horizontal lines in Figure \ref{fig:gt1} that intersect both the pc (black) and kpc (blue) curves.  In Figure \ref{fig:gt1} we show a shaded region which represents this family of lines.  The red dashed line corresponds to the best fit viewing angle in the non-bending/non-decelerating model of Section 5.1.  Thus, if we relax our non-decelerating assumption, the kpc scale jet can lie on the low $\Gamma$ tail of the K curve, without the need to invoke any jet bending.  We list the range of possible kpc scale $\Gamma$ values for this deceleration/non-bending scenario in column 10 of Table 5.  These ranges are generally narrow. In the case of the two extreme blazars, for 0106+013 we have 1.89 $< \Gamma <$ 1.92 and for 1849+670, 2.21 $< \Gamma <$ 2.23.  Thus, deceleration between pc and kpc scales offers a way to alleviate the need for unusually high bulk Lorentz factors in these blazar jets.  

\subsection{IC/CMB model with Deceleration and Jet Bending}
Many blazar jets display bent morphologies going from pc to kpc scales (e.g., \citealt{PK10}), although their apparent magnitudes are often highly exaggerated by projections effects.  For some of the jets in our sample, bending between the pc and kpc scales can lower the bulk Lorentz factor value required to reconcile the VLBI and X-ray observations, but can not change the requirement that $\Gamma \geq \ba$ on pc scales, which is derived from the superluminal motion of the radio pc scale jet.  Allowing the possibility of deceleration, acceleration, and bending effectively renders the two curves in Figure \ref{fig:gt1} independent, i.e., the kpc jet can now lie anywhere on the blue curve, and the pc jet anywhere on the black curve.  There are still, however, limits that can be placed on the beaming parameters.  For example, the expression for K (Equation \ref{eq:K_eqn}) sets an upper limit on $\theta_{kpc}$, which is a lengthy algebraic function of K \citep{HM05}.  These limits range from 8$^{\circ}$ to 20$^{\circ}$ for the jets in our sample (Figure \ref{fig:gt1}).  Setting $\mu$ = 1 for an end-on jet in Equation \ref{eq:K_eqn} also yields a lower limit of 

\begin{equation}
\Gamma_{min}=\frac{K}{2\sqrt{K-1}}
\end{equation}

\noindent on the kpc scale.  These are tabulated in column 11 of Table \ref{table:bmp}.  The IC/CMB model thus limits the kpc scale minimum bulk Lorentz factor to 1.6 $< \Gamma <$ 2.7 in most cases, although in two sources (0415+379 and 1334$-$127) the limit placed on the minimum bulk Lorentz factor must be at least 3.5.  

Finally, the superluminal speed confines the Lorentz factor of the pc jet to $\geq \ba$, and its viewing angle to below 2 $\tan^{-1}(\ba^{-1})$.  If independent observations can further constrain the amount of intrinsic jet bending, then X-ray observations of blazars can provide useful limits on the amount of pc to kpc scale jet deceleration.  It may be possible to pursue this method statistically using a larger sample.

\section{Conclusions}
We have performed Chandra observations of a radio-core-selected sample of blazar jets.  The selection criteria that we used to define our AGN sample has increased the overall fraction of correlations between X-ray and radio jets in radio selected AGN.  Of the popular single zone models available (synchrotron or IC/CMB), we chose to apply the IC/CMB model to our sample, based on the earlier results of \cite{HM05}.  The detected X-ray jets are generally well correlated spatially with the radio jet morphology, except for those radio jets that display sharp bends.  The wide range of apparent X-ray to radio ratios among the jets suggests that no single overall emission model can explain all of the X-ray morphologies.  We are currently analyzing follow-up \textit{Chandra} and \textit{HST} observations of selected AGN to obtain multiwavelength spectra of jet knots (Kharb et al. 2011, in prep.), which will allow us to investigate possible synchrotron and IC models for the emission beyond what we have discussed in this paper.

Our major findings are as follows:

\begin{itemize}
\item The selection criteria associated with the MCS has increased the detection rate from previous jet surveys \citep{RS04,HM05} from a $\sim$ 60\% detection rate to a $\sim$ 78\% detection rate.
\item We have found that the 1.4 GHz, VLA-A array extended radio jet flux density, S$_{ext}$, is a strong predictor of X-ray jet emission in a core-selected sample such as the MCS, which is related to the correlation of the extended luminosity and the pc scale jet (apparent) speed.  Above a value of 300 mJy we find a 100\% X-ray detection rate, with $\sim$ 57\% detection rate for sources located below that threshold.  This further reinforces the usefulness of our extended radio emission selection criteria for this sample.  
\item The IC/CMB assumptions can produce calculated values for the jet bulk Lorentz factor, $\Gamma$, which are larger than expected in some sources (eg. 0106+013 and 1849+670) under the assumption that the jet speed and direction are the same on both the pc and kpc scales. 
\item Bending alone can not reconcile the large $\Gamma$ values in these sources as it constrains the minimum $\Gamma$ value on the kpc scale to the minimum value on the pc scale.  This can still be quite large as seen in sources such as 0106+013 and 1849+670.
\item If we allow for the possibility of deceleration with out jet bending, the VLBI jet speeds and IC/CMB X-ray model can be reconciled, although jet bending is necessary in several cases.  In this scenario the kpc scale relativistic jet bulk Lorentz factors typically range from $\sim$ 1.7 to 7.  
\item When both the non-bending and non-decelerating assumptions are relaxed the only constraints on the kpc scale jet from the Chandra and VLA observations are an upper limit on the viewing angle, and a lower limit on the bulk Lorentz factor.  These typically range from $8^{\circ} < \theta < 20^{\circ}$ and 1.6 $< \Gamma_{min} <$ 3.5 for our sample.  
\end{itemize}

This research has made use of data from the MOJAVE database that is maintained by the MOJAVE team \citep{ML09a}.  This work was supported by Chandra Award GO8-9113A.  We would also like to thank, the referee, Dan Harris for the insightful comments he provided that helped to improve this paper.

\clearpage

\begin{figure}[b]
\begin {center}
\includegraphics [scale=.5] {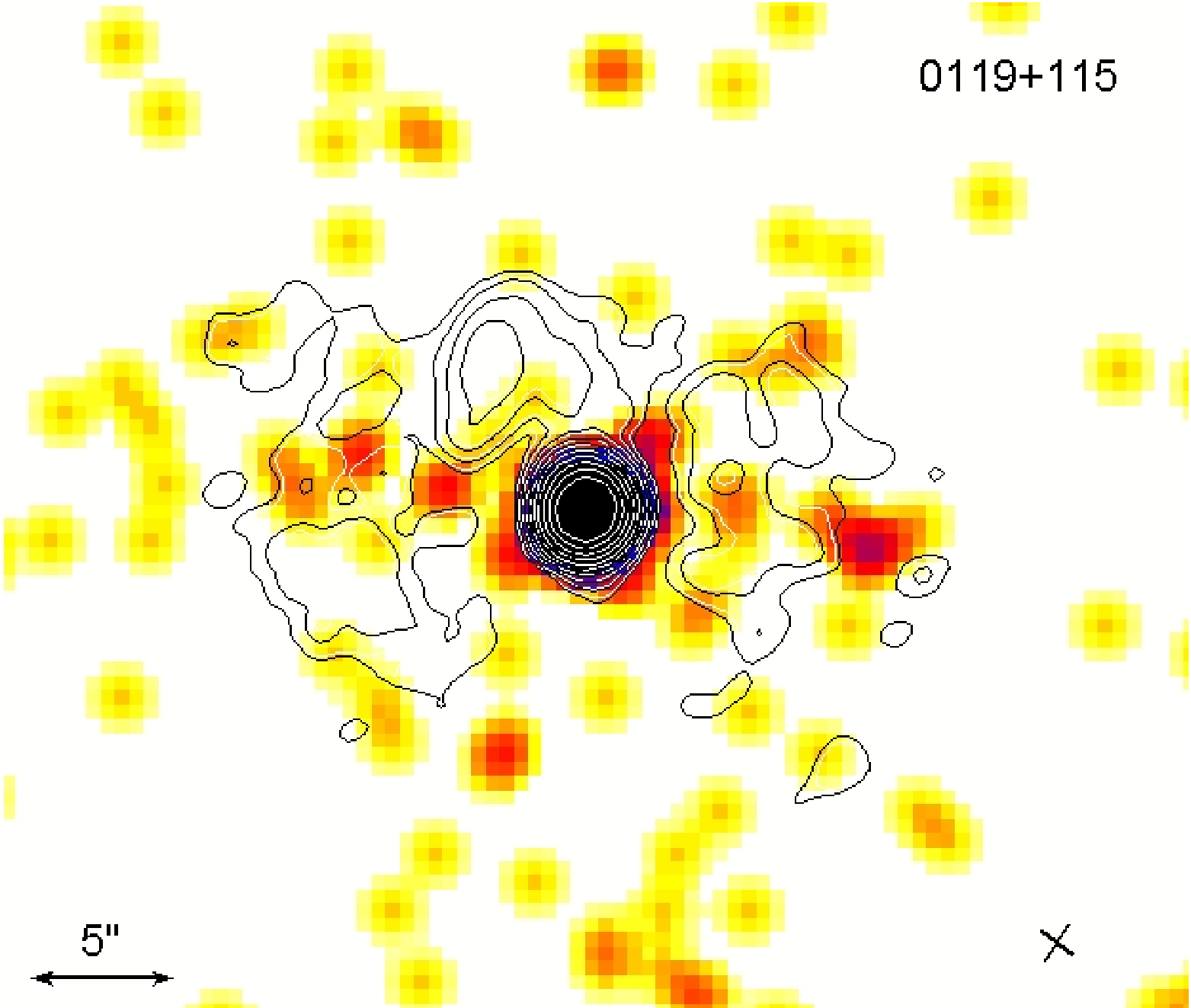}
\end{center}
\end{figure}

\begin {figure}[t]
\begin {center}
\includegraphics [scale=.5] {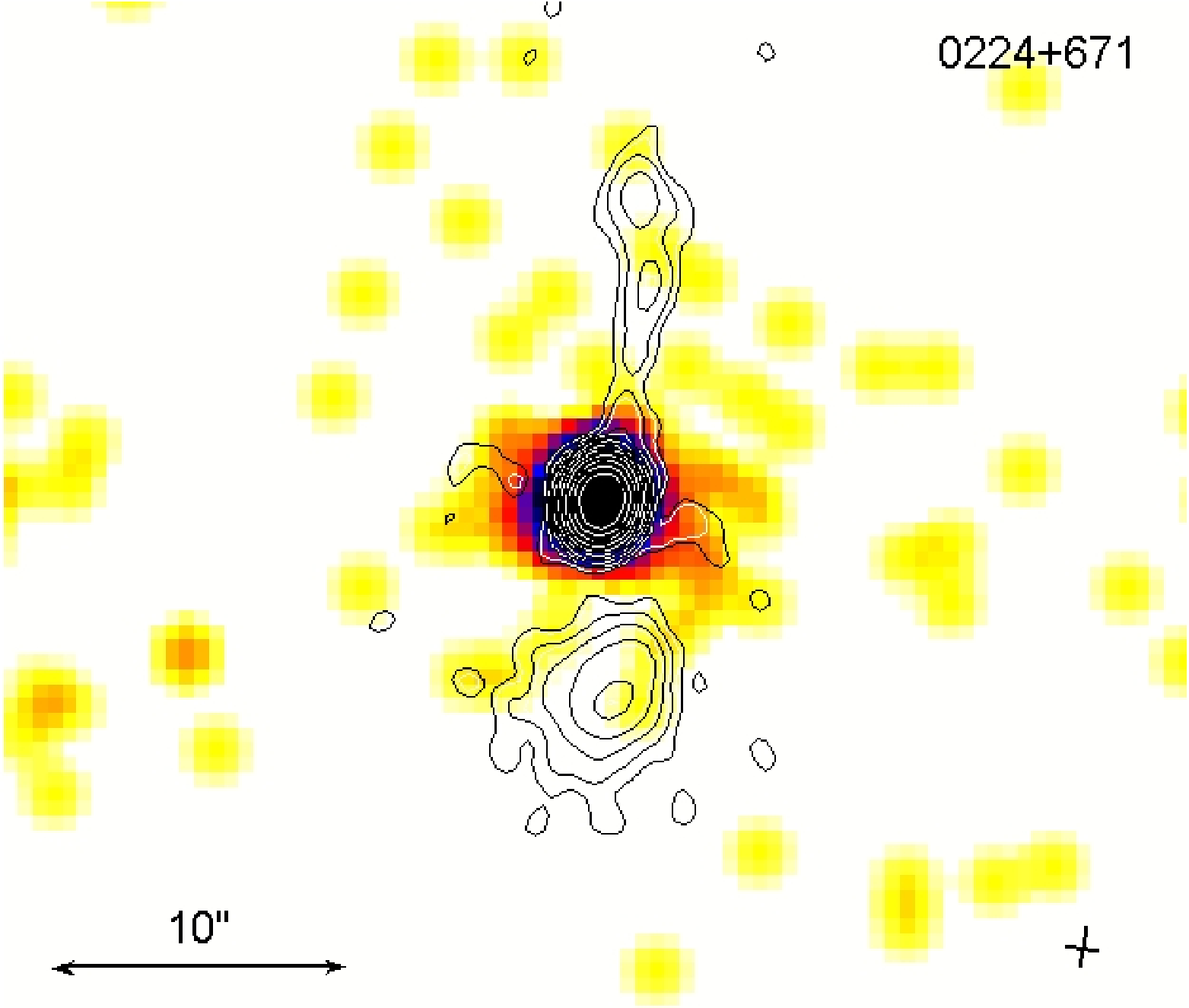}
\end{center}
\end {figure}

\clearpage

\begin {figure}[b]
\begin {center}
\includegraphics [scale=.5] {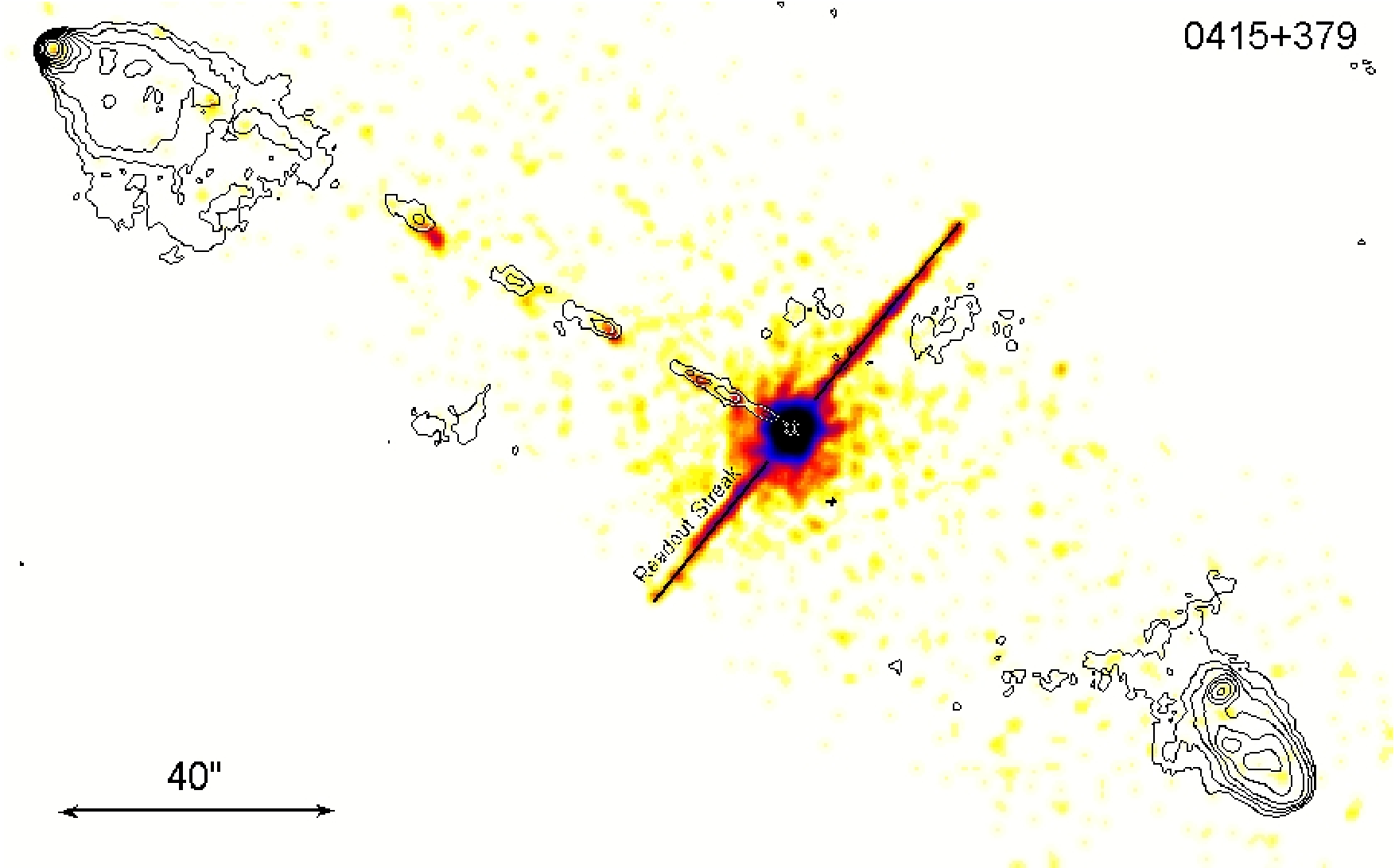}
\end{center}
\end{figure}

\clearpage

\begin {figure}[t]
\begin {center}
\includegraphics [scale=.5] {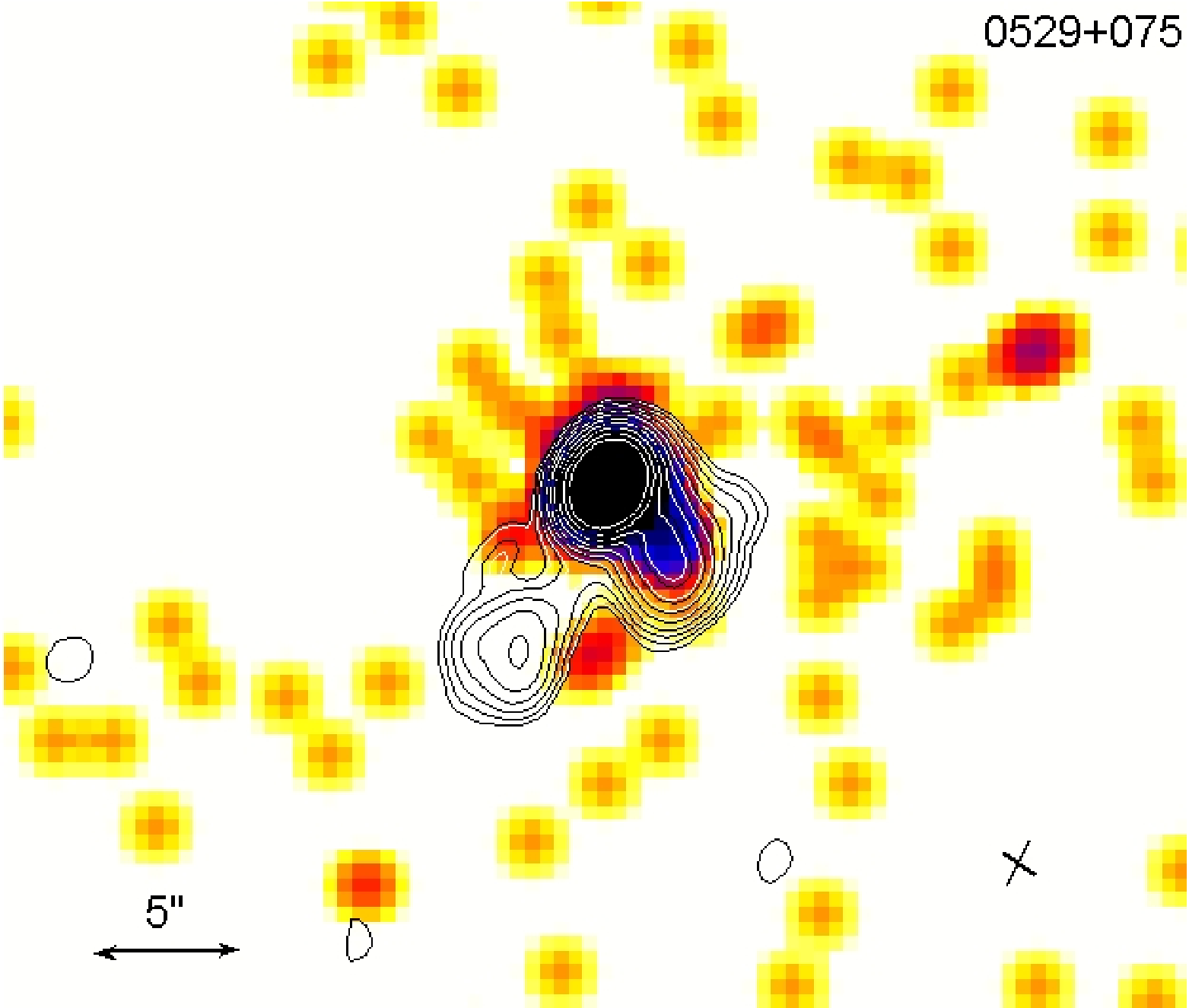}
\end{center}
\end {figure}

\begin {figure}[b]
\begin {center}
\includegraphics [scale=.5] {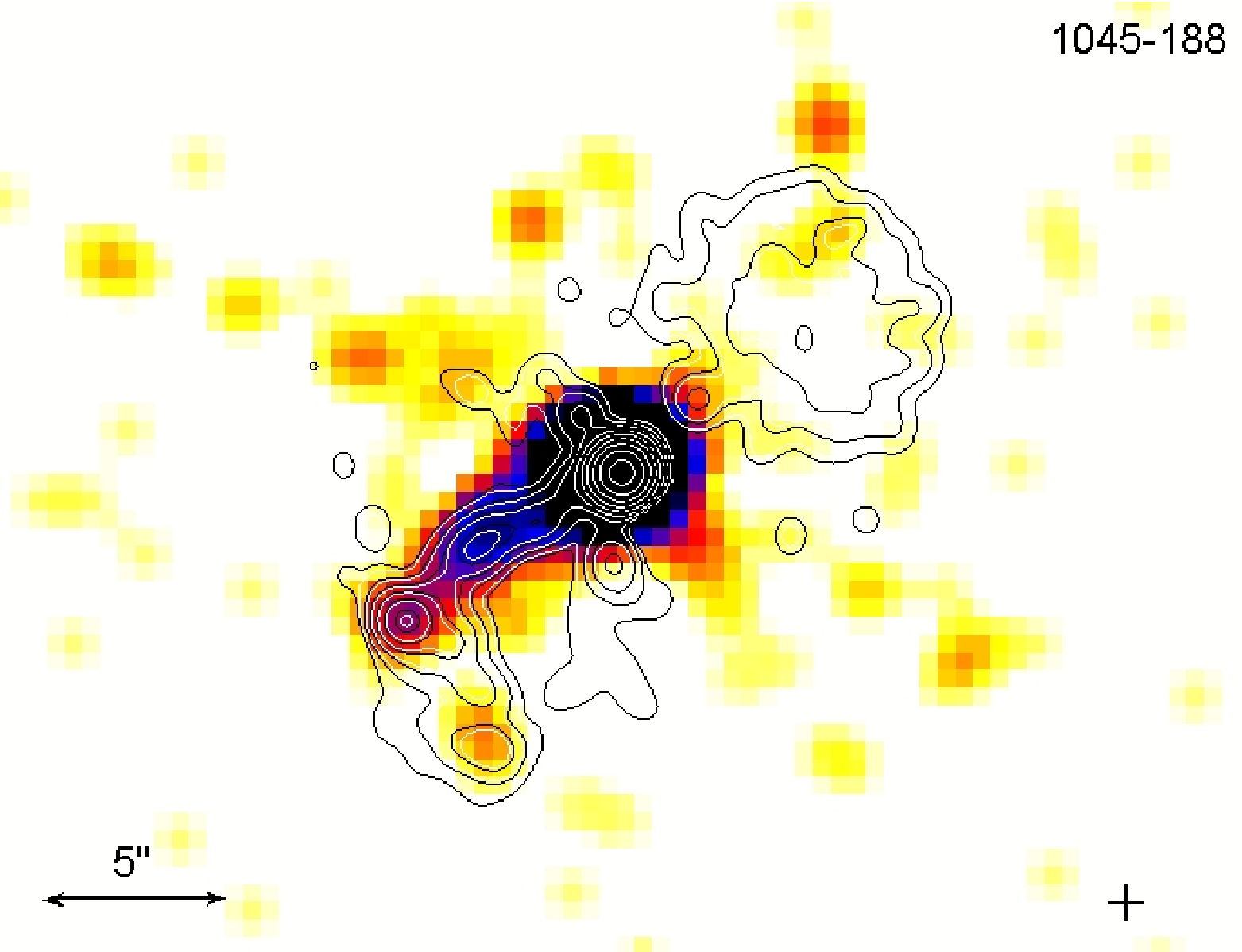}
\end{center}
\end{figure}

\clearpage

\begin {figure}[t]
\begin {center}
\includegraphics [scale=.5] {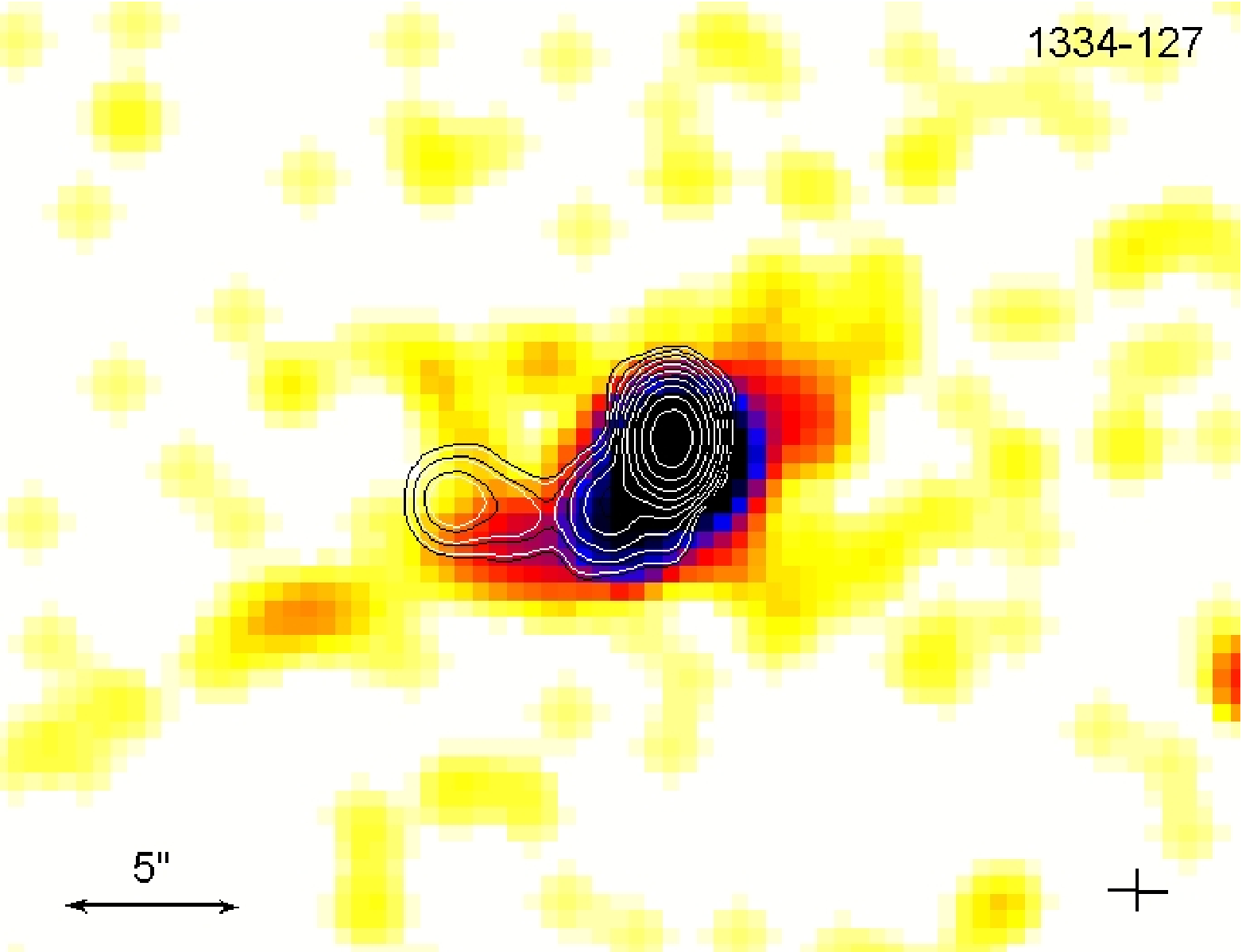}
\end{center}
\end {figure}

\begin {figure}[b]
\begin {center}
\includegraphics [scale=.5] {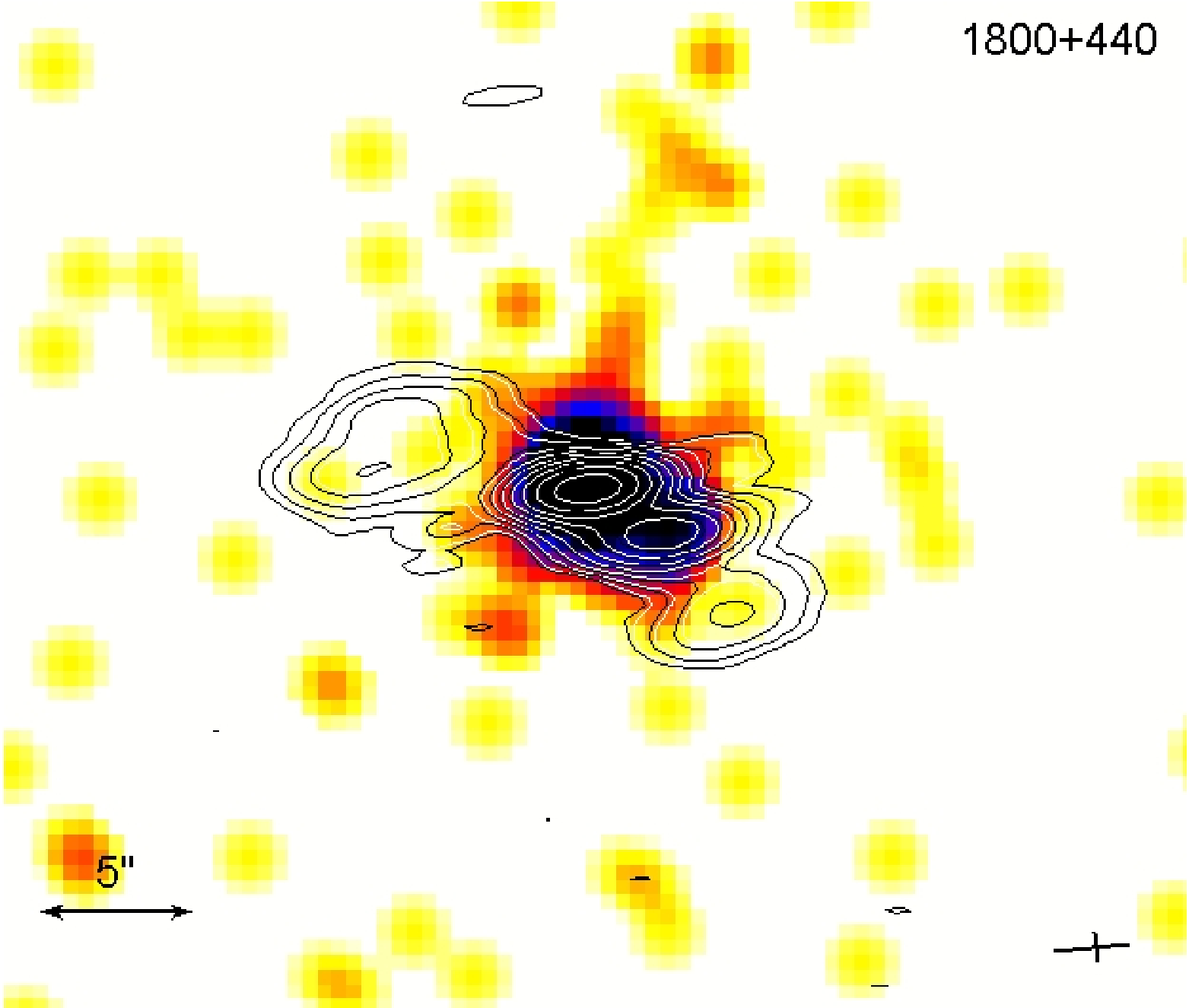}
\end{center}
\end {figure}

\clearpage

\begin {figure}[t]
\begin {center}
\includegraphics [scale=.5] {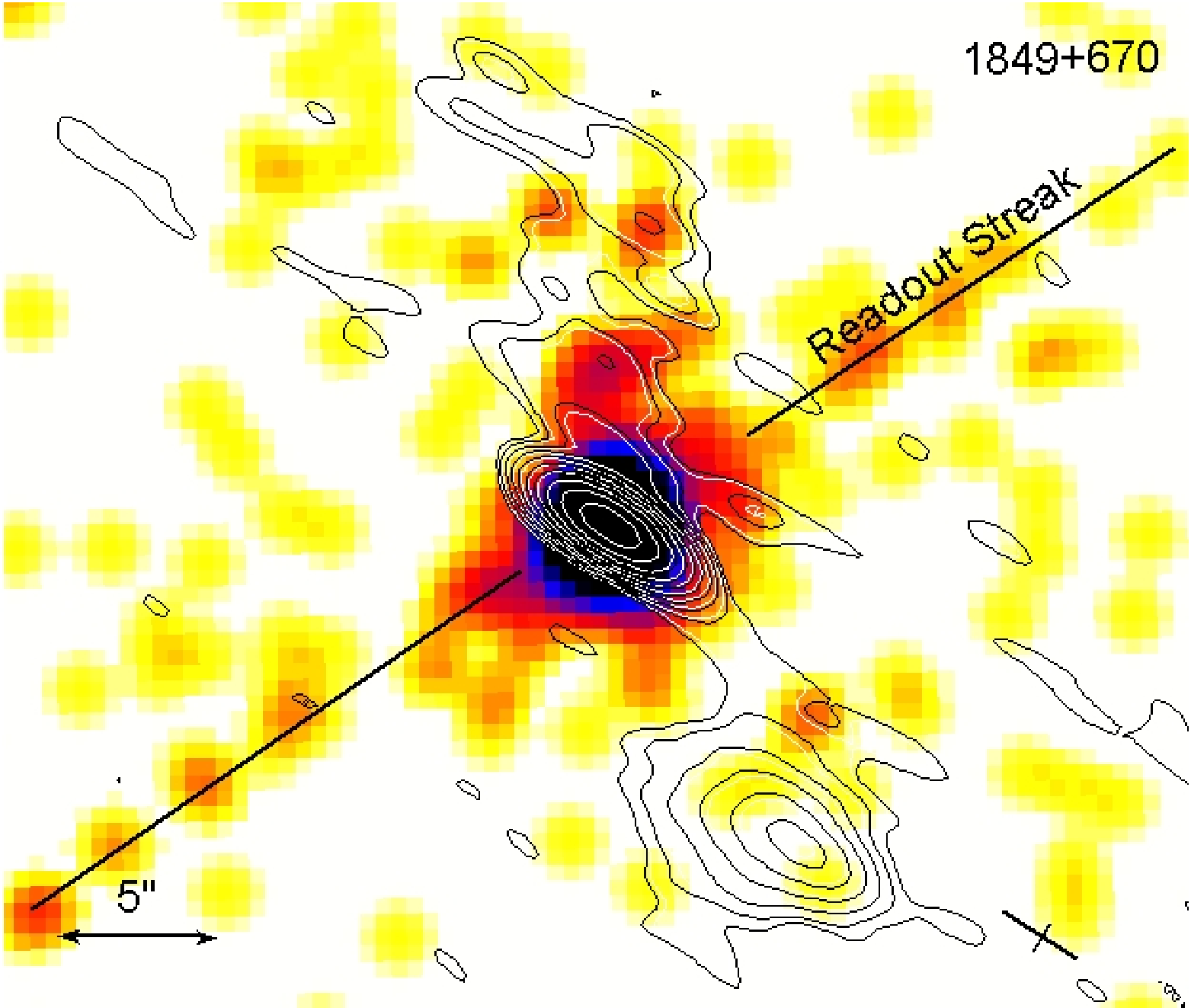}
\end{center}
\end{figure}

\begin {figure}[b]
\begin {center}
\includegraphics [scale=.5] {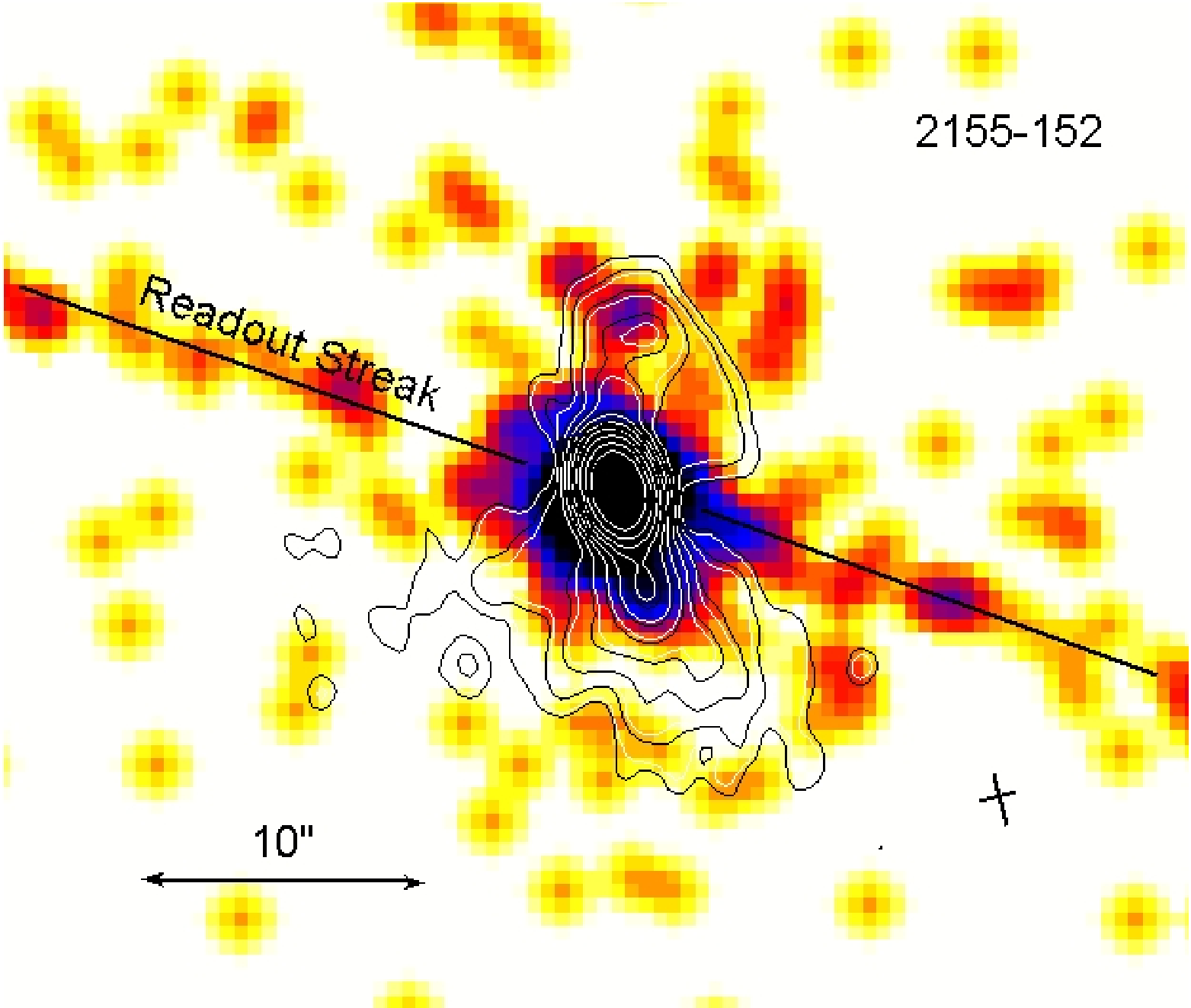}
\end{center}
\end {figure}

\clearpage

\begin {figure}[t]
\begin {center}
\includegraphics [scale=7.0] {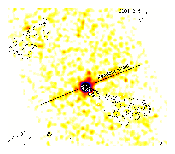}
\end{center}
\end{figure}

\begin {figure}[b]
\begin {center}
\includegraphics [scale=10] {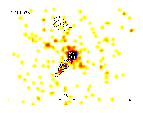}
\end{center}
\end {figure}

\clearpage

\begin {figure}[t]
\begin {center}
\includegraphics [scale=.5] {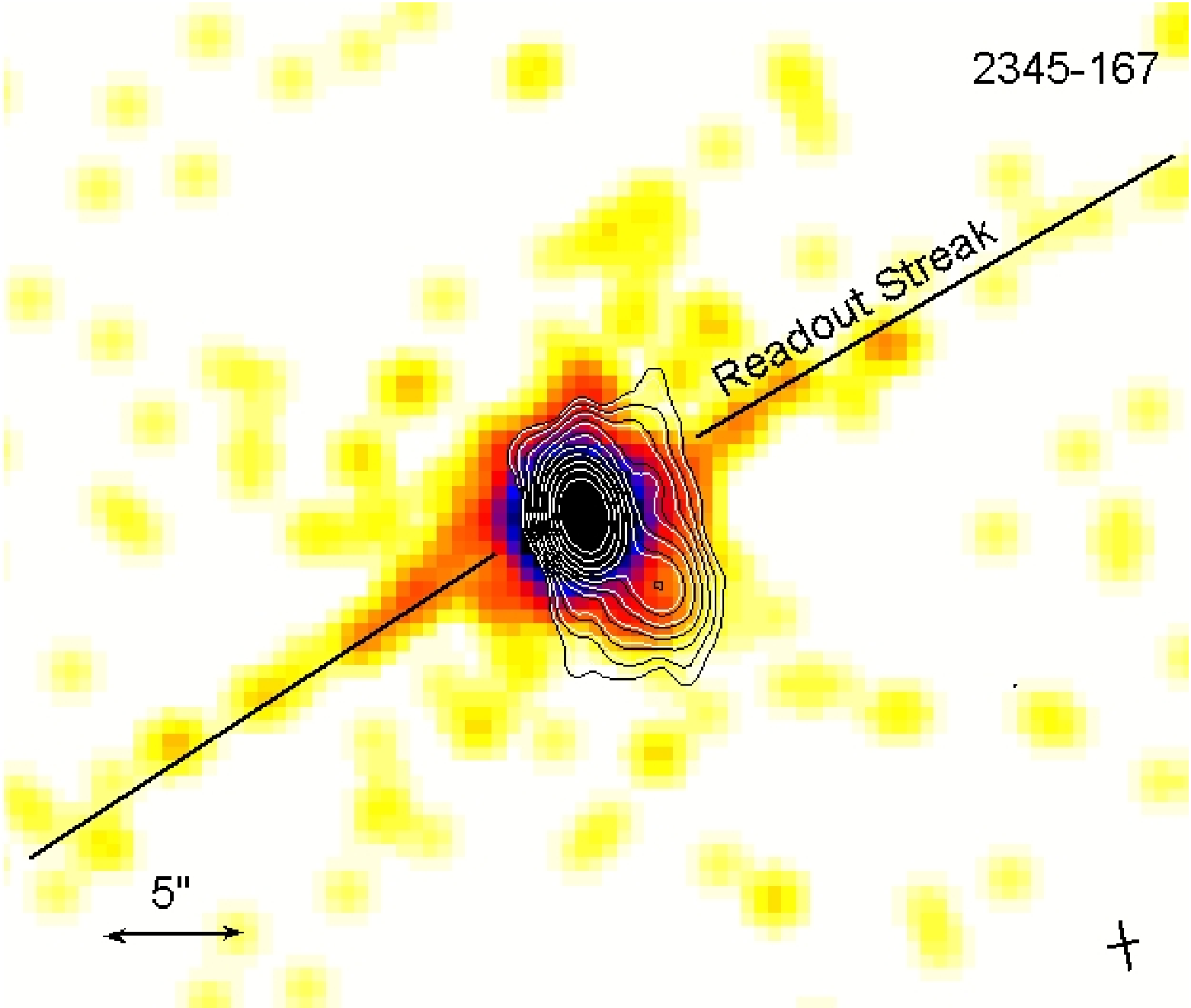}
\end{center}
\end{figure}

\begin{figure}
\includegraphics {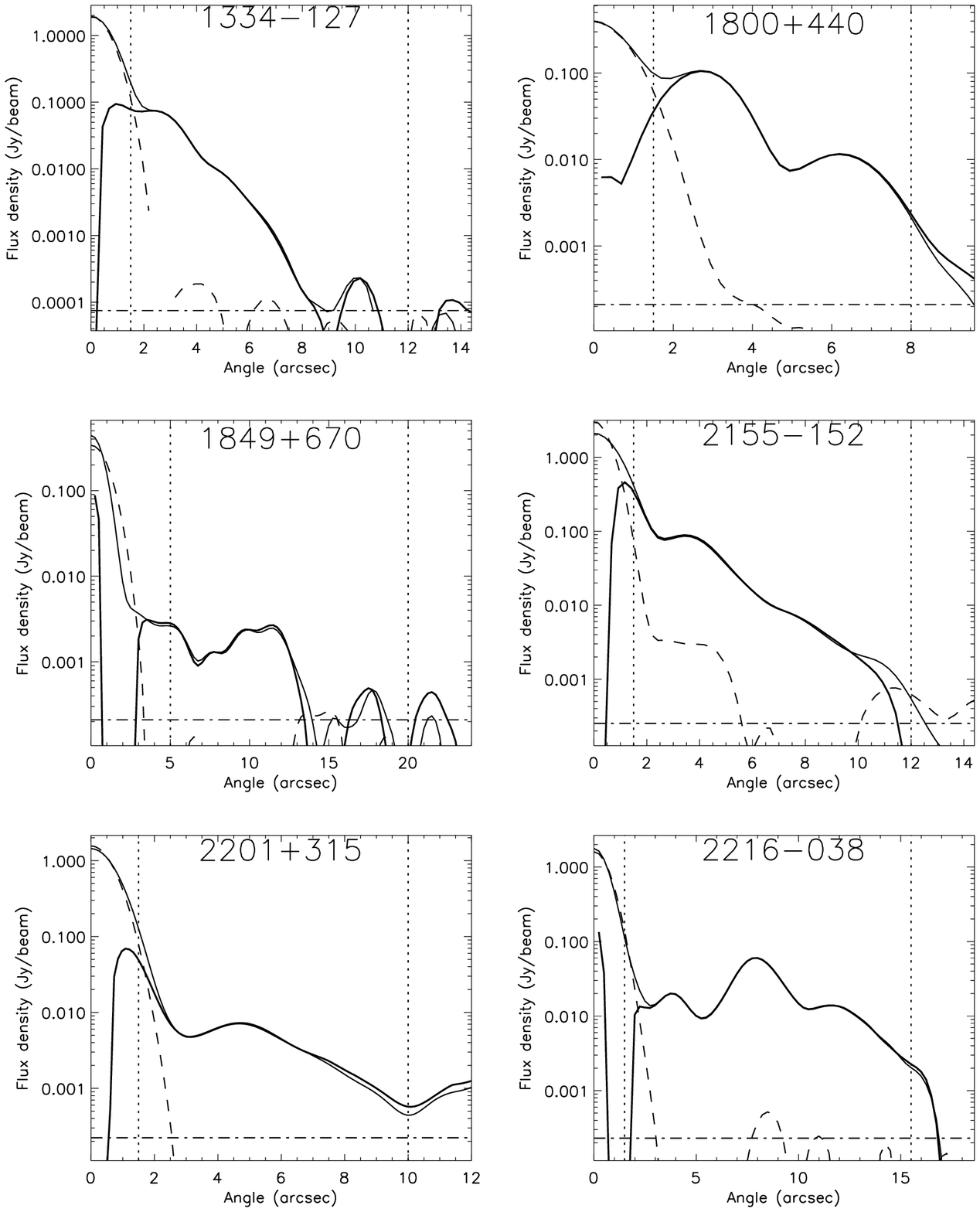}
\end{figure}

\begin{figure}
\includegraphics {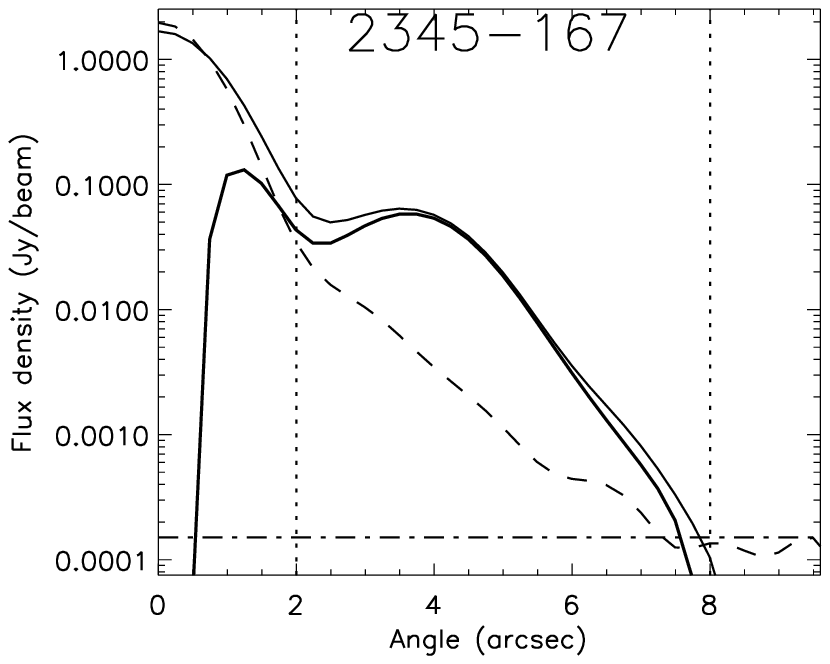}
\end{figure}

\begin{figure}
\includegraphics {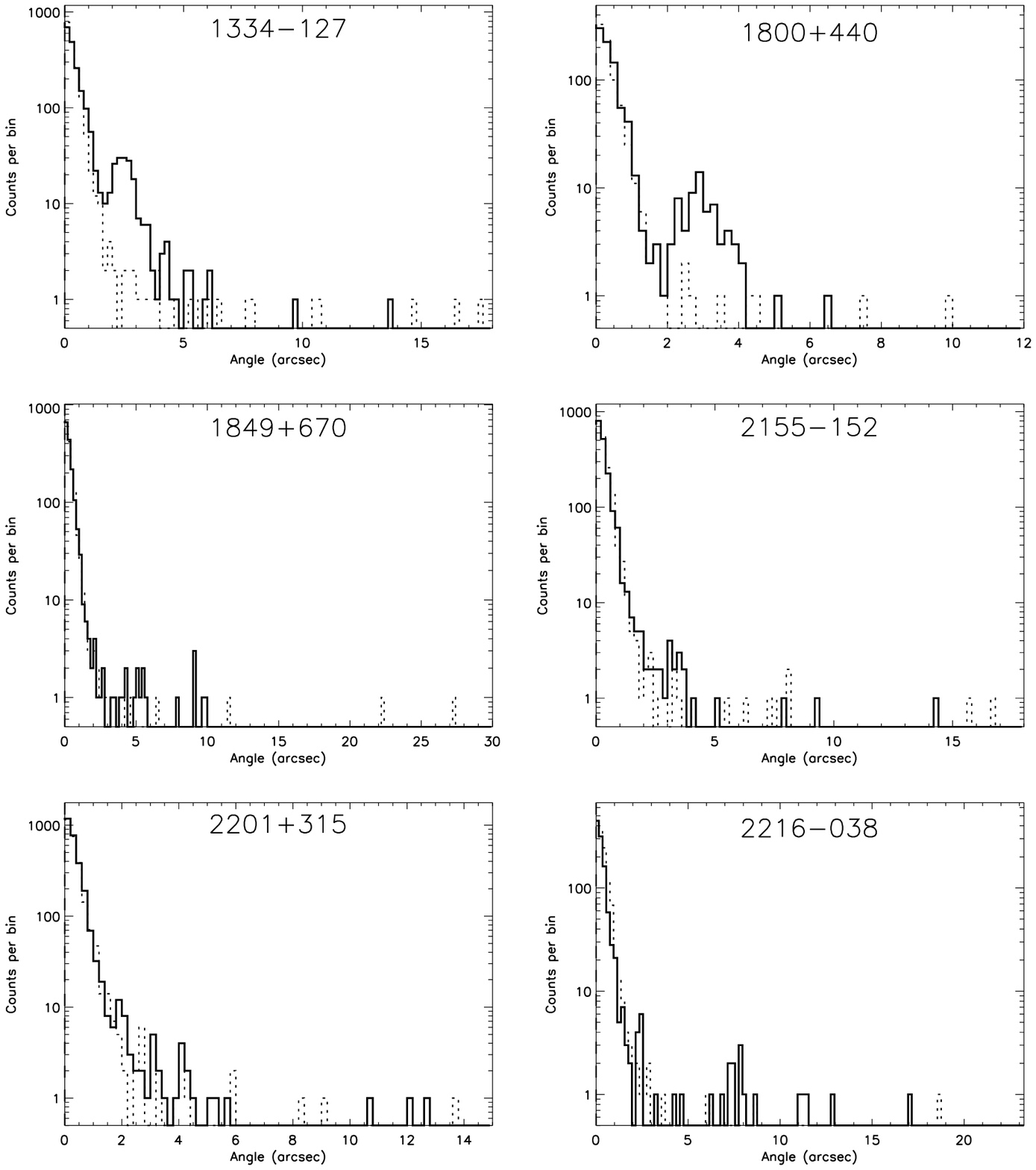}
\end{figure}

\begin{figure}
\includegraphics {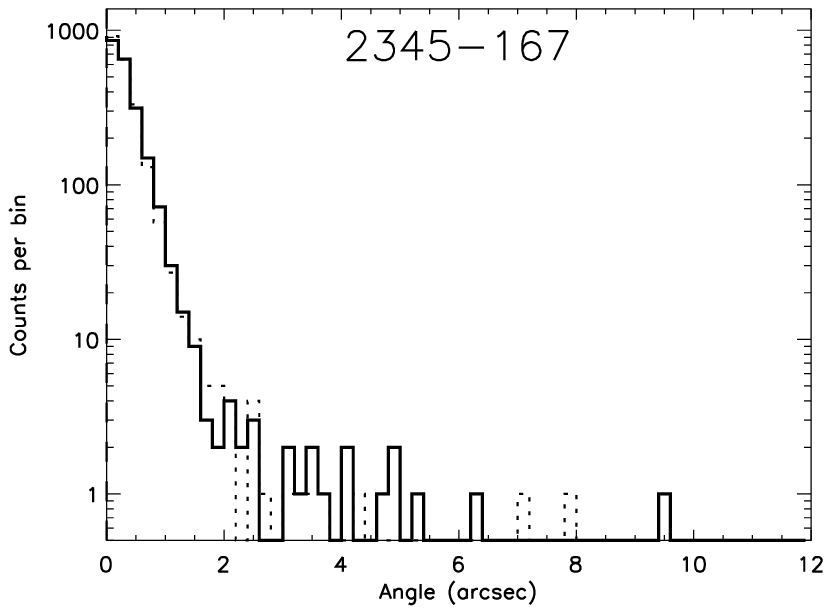}
\end{figure}

\begin{figure}
\includegraphics [scale=.4]{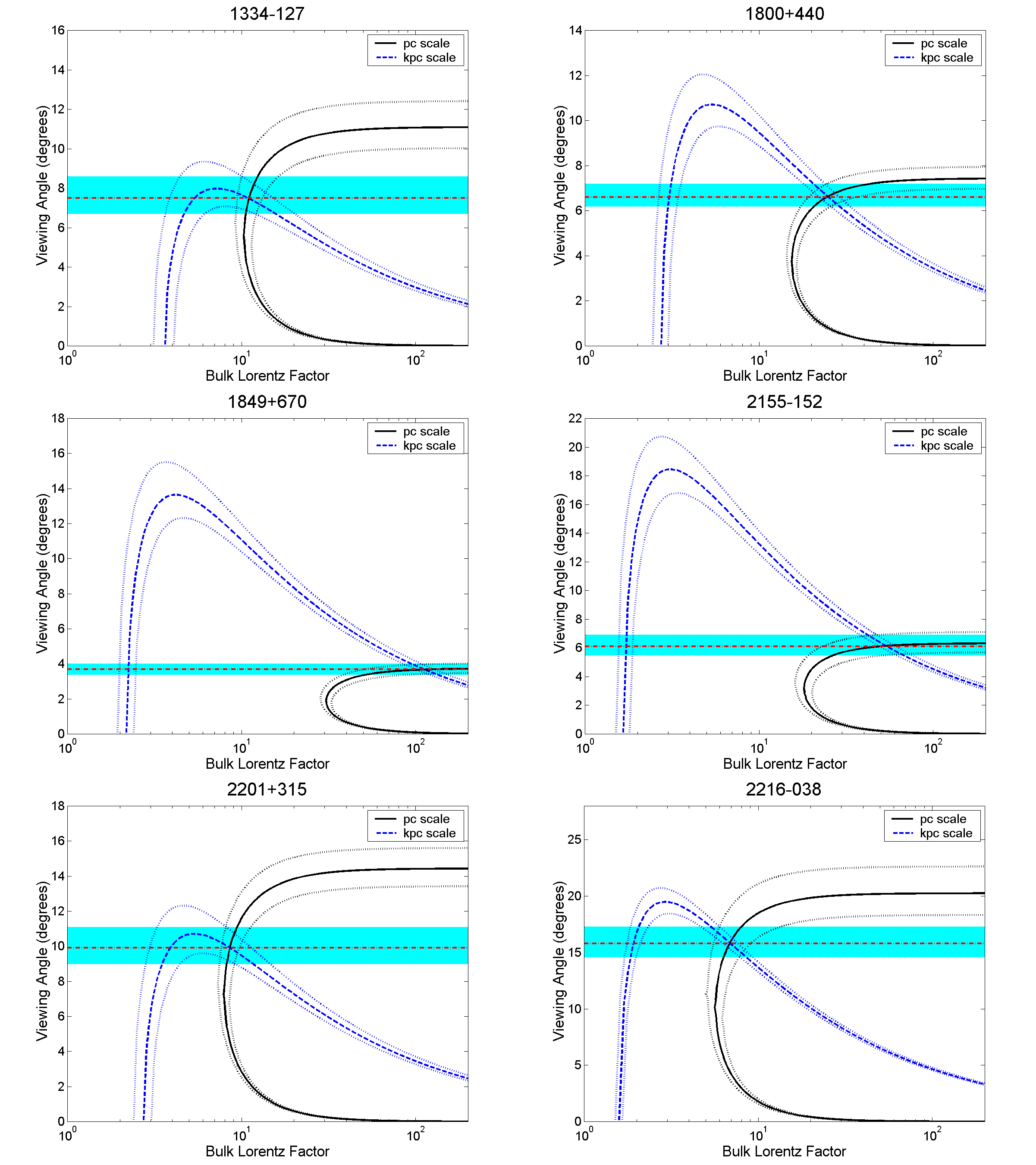}
\end{figure}

\begin{deluxetable}{lllllll}
 \tabletypesize{\small}
\tablecolumns{7}
\tablecaption{\label{parameters}MOJAVE CHANDRA SAMPLE}
\tablewidth{0pt}
\tablehead{\colhead{Source} & \colhead {Alias} & \colhead{z} & \colhead{S$_{ext}$} & \colhead{$\beta$$_{app}$} &\colhead{Reference} & \colhead {Obs ID }\\
\colhead{(1)} & \colhead{(2)} & \colhead{(3)} & \colhead{(4)} & \colhead{(5)} & \colhead{(6)} & \colhead{(7)}}
\vline
\startdata
0106$+$013& OC 12 				& 2.099 		& 0.53  	& 26.5  $\pm$   4.2        		& This paper & 9281 \\
0119$+$115&								& 0.57  		& 0.11   	& 17.1	$\pm$		1.1						& This paper& 9290 \\
0224$+$671& 4C 67.05			& 0.523 		& 0.15 		& 11.6 	$\pm$		0.8						& This paper& 9288 \\
0234$+$285& CTD 20				& 1.207 		& 0.10 		& 12.3 	$\pm$		1.1						& Marshall et al. 2005 & 4898 \\
0415$+$379& 3C 111 				& 0.0491 		& 2.70		& 5.9 	$\pm$		0.3						& This paper&	9279 \\
0529$+$075& OG 050 				& 1.254 		& 0.13 		& 12.7 	$\pm$		1.6						& This paper & 9289 \\
0605$-$085& 	  					& 0.872 		& 0.12 		& 19.8 	$\pm$		1.2						& Sambruna et al. 2004 & 2132 \\
1045$-$188& 	 					  & 0.595 		& 0.51 		& 8.6		$\pm$ 	0.8						& This paper & 9280 \\
1055$+$018& 4C 01.28			& 0.89  		& 0.23 		& 11.0 	$\pm$		1.2						& Sambruna et al. 2004& 2137 \\
1156$+$295& 4C 29.45			& 0.729 		& 0.20 		& 24.9 	$\pm$		2.3						& Coppi et al. & 0874 \\
1222$+$216& 4C 21.35 			& 0.432 		& 0.96 		& 21.0 	$\pm$		2.2						& Jorstad \& Marscher 2005 & 3049 \\
1226$+$023& 3C 273				& 0.158 		& 17.67   & 13.4 	$\pm$		0.8						& Jester et al. 2006 & 4879 \\
1253$-$055& 3C 279				& 0.536 		& 2.10 		& 20.6 	$\pm$	 	1.4						& WEBT Collaboration 2007 & 6867 \\
1334$-$127& 		 					& 0.539 		& 0.15 		& 10.3 	$\pm$		1.1						& This paper& 9282 \\
1510$-$089& 							& 0.36 			& 0.18 		& 20.2 	$\pm$		4.9						& Sambruna et al. 2004& 2141 \\
1641$+$399& 3C 345 				& 0.593 		& 1.48 		& 19.3 	$\pm$		1.2						& Sambruna et al. 2004& 2143 \\
1655$+$077& 							& 0.621 		& 0.20 		& 14.4 	$\pm$		1.4						& Marshall et al. 2005& 3122 \\
1800$+$440& S4 1800$-$44 	& 0.663 		& 0.25 		& 15.4 	$\pm$		1.0						& This paper & 9286 \\
1828$+$487& 3C 380 				& 0.692 		& 5.43 		& 13.7 	$\pm$		0.8						& Marshall et al. 2005& 3124 \\
1849$+$670& S4 1849$-$67 	& 0.657  		& 0.10 		& 30.6 	$\pm$		2.2						& This paper & 9291 \\
1928$+$738& 4C 73.18			& 0.302 		& 0.36 		& 8.4	  $\pm$		0.6						& Sambruna et al. 2004 & 2145 \\
1957$+$405& Cygnus A 			& 0.0561 		& 414.18 	& 0.2 	$\pm$		0.1						& Wilson et al. 2001 & 1707 \\
2155$-$152& 							& 0.672 		& 0.30 		& 18.1 	$\pm$		2.0						& This paper & 9284 \\
2201$+$315& 4C 31.63			& 0.295 		& 0.37 		& 7.9 	$\pm$		0.6						& This paper & 9283 \\
2216$-$038& 							& 0.901 		& 0.31 		& 5.6 	$\pm$		0.6						& This paper & 9285 \\
2251$+$158& 3C 454.3			& 0.859 		& 0.88 		& 14.2 	$\pm$		1.1						& Marshall et al. 2005& 3127 \\
2345$-$167& 							& 0.576 		& 0.14 		& 13.5 	$\pm$		1.1						& This paper & 9328 \\
\enddata 
\tablecomments{\tiny{Columns are as follows:\\
(1)	IAU name (B1950.0)\\
(2)	Common Name\\
(3)	Redshift from NED\\
(4)	Extended flux density (total  - core) at 1.4 GHz (Jy)\\
(5)	Superluminal velocity in units of the speed of light \citep{ML09b}\\
(6)	Reference for X-ray image\\
(7)	\textit{Chandra} observation ID number}}
\label{table:mcs}
\end{deluxetable}

\begin{deluxetable}{ccccc}
\tabletypesize{\scriptsize}
\tablecolumns{5}
\tablecaption{\label{parameters}OBSERVATION LOG}
\tablewidth{0pt}
\tablehead{\colhead{Source}  & \colhead{Live Time} & \colhead{Date} & \colhead {RA} & \colhead {DEC}\\
\colhead{(1)} & \colhead{(2)} & \colhead{(3)} & \colhead{(4)} & \colhead{(5)}} 
 
\vline
\startdata
0106$+$013  & 9.69  & 2007 Nov 21 & 1h8m38.771s & +1$^\circ$35$\arcmin$0.317$\arcsec$\\
0119$+$115  & 9.95  & 2008 Oct 27 & 1h21m41.595s & +11$^\circ$49$\arcmin$50.413$\arcsec$\\ 
0224$+$671  & 10.11 & 2008 Jun 27 & 2h28m50.051s & +67$^\circ$21$\arcmin$3.029$\arcsec$\\
0415$+$379  & 10.14 & 2008 Dec 10 & 4h18m21.277s & +38$^\circ$1$\arcmin$35.800$\arcsec$\\ 
0529$+$075  & 10.18 & 2007 Nov 16 & 5h32m38.998s & +7$^\circ$32$\arcmin$43.345$\arcsec$\\  
1045$-$188  & 10.18 & 2008 Apr 01 & 10h48m6.621s & $-$19$^\circ$9$\arcmin$35.727$\arcsec$\\
1334$-$127  & 10.79 & 2008 Mar 09 & 13h37m39.783s& $-$12$^\circ$57$\arcmin$24.693$\arcsec$\\
1800$+$440  & 10.19 & 2008 Jan 05 & 18h1m32.315s & +44$^\circ$4$\arcmin$21.900$\arcsec$\\
1849$+$670  & 10.19 & 2008 Feb 27 & 18h49m16.072s & +67$^\circ$5$\arcmin$41.680$\arcsec$\\ 
2155$-$152  & 10.19 & 2008 Jul 10 & 21h58m6.282s & $-$15$^\circ$1$\arcmin$9.328$\arcsec$\\
2201$+$315  & 10.11 & 2008 Oct 12 & 22h3m14.976s & +31$^\circ$45$\arcmin$38.270$\arcsec$\\
2216$-$038  & 10.16 & 2007 Dec 02 & 22h18m52.038s & $-$3$^\circ$35$\arcmin$36.879$\arcsec$\\
2345$-$167  & 10.15 & 2008 Sep 01 & 23h48m2.609s & $-$16$^\circ$31$\arcmin$12.022$\arcsec$\\ 
 
\enddata 
\tablecomments{\tiny{Columns are as follows:\\
(1) IAU name (B1950.0)\\
(2)	Exposure time in kiloseconds\\
(3) Date observed\\
(4) Right ascension of the radio core position from NED(J2000)\\
(5) Declination of the radio core position from NED (J2000)\\}}
\label{table:obs}
\end{deluxetable}

 \begin{deluxetable}{ccccccc}
\tabletypesize{\scriptsize}
\tablecolumns{7}
\tablecaption{\label{parameters}VLA Archival Data}
\tablewidth{0pt}
\tablehead{\colhead{Source}  & \colhead{Observation Date} & \colhead{Project} & \colhead{RMS Noise} & \colhead{B$_{maj}$} & \colhead{B$_{min}$} & \colhead{B$_{maj}$ PA} \\
\colhead{(1)} & \colhead{(2)} & \colhead{(3)} & \colhead{(4)} & \colhead{(5)}& \colhead{(6)} & \colhead{(7)} }

\startdata
0106$+$013   & 2004-09-19 & AL634 & 1.4$\times$10$^{-1}$ & 1.64  	& 1.49	& 100\\
0119$+$115   & 2004-09-19 & AL634 & 9.8$\times$10$^{-2}$ & 1.53  	& 1.44	& 55\\
0224$+$671   & 2004-09-19 & AL634 & 1.5$\times$10$^{-1}$ & 1.42  	& 1.13	& 78\\
0415$+$379   & 1982-06-14 & LINF  & 1.9$\times$10$^{-1}$ & 1.60  	& 1.47	& 168\\
0529$+$075   & 2004-09-19 & AL634 & 4.6$\times$10$^{-2}$ & 1.70    & 1.35	& 63\\
1045$-$188   & 2007-06-30 & AC874 & 3.4$\times$10$^{-1}$ & 1.00  	& 1.00	& 90\\
1334$-$127   & 1986-03-18 & AD176 & 7.1$\times$10$^{-2}$ & 1.73  	& 1.22	& 89\\
1800$+$440   & 1990-05-18 & AS396 & 1.8$\times$10$^{-1}$ & 2.54  	& 1.02	& 7\\
1849$+$670   & 2004-11-09 & AL634 & 2.0$\times$10$^{-1}$ & 2.77  	& 1.06	& 146\\
2155$-$152   & 2004-11-21 & AL634 & 2.0$\times$10$^{-1}$ & 1.90  	& 1.26	& 106\\
2201$+$315   & 2004-11-21 & AL634 & 1.1$\times$10$^{-1}$ & 1.57  	& 1.43	& 164\\
2216$-$038   & 2004-11-21 & AL634 & 1.8$\times$10$^{-1}$ & 1.58  	& 1.34	& 113\\
2345$-$167   & 2004-11-09 & AL634 & 1.6$\times$10$^{-1}$ & 1.88  	& 1.22	& 102\\
\enddata

\tablecomments{\tiny{Columns are as follows:\\
(1) IAU name (B1950.0)\\
(2) Date observed\\
(3) Project code\\
(4) Rms noise level of radio image in mJy beam$^{-1}$\\
(5) Major axis for the radio beam in ($\arcsec$)\\
(6) Minor axis for the radio beam in ($\arcsec$)\\
(7) Position angle of the radio beam major axis in ($^\circ$)\\}}
\label{table:vla}
\end{deluxetable}

\begin{deluxetable}{rrrrrrrrrrc}
\tabletypesize{\scriptsize}
\tablecolumns{11}
\tablecaption{\label{parameters}MOJAVE CHANDRA SAMPLE: JET MEASUREMENTS}
%\rotate
\tablewidth{0pt}
\tablehead{\colhead{Source}  & \colhead{$PA_{pc}$} & \colhead{$PA_{kpc}$} & \colhead{r{$_i$}} & \colhead{r{$_o$}} & \colhead{S{$_ r$}} & \colhead{$\nu${$_r$}} &\colhead{Count Rate} & \colhead{S{$_x$}}  & \colhead{P{$_{jet}$}} &\colhead{X-Jet}\\
\colhead{(1)} & \colhead{(2)} & \colhead{(3)} & \colhead{(4)} & \colhead{(5)} & \colhead{(6)} & \colhead{(7)} & \colhead{(8)}& \colhead{(9)} & \colhead{(10)} & \colhead{(11)}}

\startdata
0106$+$013 & $-$122 &  180   &  1.5 &  8.0 &   526.7 $\pm$ 0.4 &  1.40 &   9.90 $\pm$  1.11   &     9.9  &$<$ 1$\times10^{-10}$ & Y \\
0119$+$115 & 6      &   35   &  1.5 &  8.0 &    22.3 $\pm$ 0.3 &  1.40 &   0.00 $\pm$  0.38   & $<$ 1.2  &5.54$\times10^{-1}$ & N \\
0224$+$671 & $-$5   &  $-$10 &  1.5 & 11.0 &    22.9 $\pm$ 0.8 &  1.40 &  $-$0.55 $\pm$  0.45 & $<$ 0.8  &9.62$\times10^{-1}$ & N \\
0415$+$379 & 71     &   63   &  1.5 & 100.0&    50.6 $\pm$ 7.2 &  1.44 &   7.50 $\pm$  2.49   &     7.5  &2.44$\times10^{-6}$ & Y \\
0529$+$075 & $-$7   & $-$145 &  1.5 &  8.0 &    69.2 $\pm$ 0.3 &  1.40 &   1.52 $\pm$  0.65   &     1.5  &1.99$\times10^{-4}$ & Y \\
1045$-$188 & 146    &  125   &  1.5 & 10.0 &   167.8 $\pm$ 5.0 &  1.42 &   2.82 $\pm$  0.82   &     2.8  &4.74$\times10^{-8}$ & Y \\
1334$-$127 & 147    &  135   &  1.5 & 12.0 &   103.9 $\pm$ 0.3 &  1.49 &  17.07 $\pm$  1.56   &    17.1  &$<$ 1$\times10^{-10}$ & Y \\
1800$+$440 & $-$157 & $-$130 &  1.5 &  8.0 &   133.2 $\pm$ 0.5 &  1.51 &   6.28 $\pm$  0.99   &     6.3  &$<$ 1$\times10^{-10}$ & Y \\
1849$+$670 & $-$52  &    0   &  5.0 & 20.0 &     8.3 $\pm$ 0.8 &  1.40 &   1.08 $\pm$  0.40   &     1.1  &1.36$\times10^{-6}$ & Y \\
2155$-$152 & $-$146 & $-$170 &  1.5 & 12.0 &   231.4 $\pm$ 0.9 &  1.40 &   1.51 $\pm$  0.85   &     1.5  &5.00$\times10^{-3}$ & Y \\
2201$+$315 & $-$141 & $-$110 &  1.5 & 10.0 &    31.1 $\pm$ 0.7 &  1.40 &   1.96 $\pm$  1.05   &     2.0  &1.54$\times10^{-3}$ & Y \\
2216$-$038 & $-$170 &  135   &  1.5 & 15.5 &   164.3 $\pm$ 1.0 &  1.40 &   1.74 $\pm$  0.78   &     1.7  &4.94$\times10^{-4}$ & Y \\
2345$-$167 & 141    & $-$135 &  2.0 &  8.0 &    83.8 $\pm$ 0.4 &  1.40 &   0.65 $\pm$  0.67   & $<$ 2.7  &8.92$\times10^{-2}$ & N \\
\enddata 

\tablecomments{\tiny{Columns are as follows:\\
(1)	IAU name (B1950.0)\\
(2)	Position angle of the pc-scale radio jet in ($^\circ$). All position angles are measured from north through east.\\ 
(3)	Position angle of the kpc-scale radio jet ($^\circ$)\\ 
(4)	Inner radius in ($\arcsec$) (see Section 2.2)\\
(5)	Outer radius in ($\arcsec$) (see Section 2.2)\\
(6) Observed flux density of the radio jet in mJy, measured in the region defined by the same region as for the X-ray count rate, given by the $PA_{kpc}$, R$_{i}$, and R$_{o}$ parameters.\\
(7) Observation frequency of the radio image in GHz.\\
(8) Counts per kilosecond in the region defined by the $PA_{kpc}$, R$_{i}$, and R$_{o}$ parameters. (see section 2.2)\\
(9) The X-ray flux density (nJy) is given at 1 keV assuming a conversion of 1 Jy s Count$^{-1}$, which is good to 10$\%$ for power law spectra with low column densities and spectral indices ($\alpha_{x}$) near 1.5.\\
(10) Probability of having more counts than those observed in the specified region under the null hypothesis that the counts are background events.\\
(11) X-ray jet detection.  The jet is defined to be detected if P$_{jet}$ $<$ 0.0025 (see Section 2.2).}}
\label{table:sjm}
\end{deluxetable}

\begin{deluxetable}{rrrrlrrcrrcc}
\tabletypesize{\scriptsize}
\tablecolumns{12}
\tablecaption{\label{parameters}MOJAVE CHANDRA SAMPLE: JET PROPERTIES}
\tablewidth{0pt}
\tablehead{\colhead{Source}  & \colhead{R} & \colhead{V} & \colhead{$B_{1}$} & \colhead{K} & \colhead{$\alpha${$_{rx}$}} & \colhead{$\theta$} & \colhead{$r_{o,deproj}$}  & \colhead{$\delta$} & \colhead{$\Gamma_{pc=kpc}$} & \colhead{$\Gamma_{kpc,decel}$}  & \colhead{$\Gamma_{kpc,min}$}\\
\colhead{(1)} & \colhead{(2)} & \colhead{(3)} & \colhead{(4)} & \colhead{(5)} & \colhead{(6)} & \colhead{(7)} & \colhead{(8)}& \colhead{(9)} & \colhead{(10)} & \colhead{(11)}& \colhead{(12)}}

\startdata
0106$+$013 &        0.0740 & 1.7$\times$10$^{3}$ &  146 &      13$\pm$2   &     0.94 $\pm$ 0.01		& 4.2	  & 922 & 3.6	& $99^{+33}_{-29}$ & 1.89$-$1.92 & 1.9\\
0119$+$115 &   $<$  0.2010 & 8.0$\times$10$^{2}$ &   29 & $<$  19         & $>$ 0.88				      & $>$ 6.3	  & ... & 4.4	& 36 & ...  & ...\\
0224$+$671 &   $<$  0.1322 & 1.0$\times$10$^{3}$ &   26 & $<$  14         & $>$ 0.91				      & $>$ 8.9	  & ... & 3.8	& 20 & ...  & ...\\
0415$+$379 &        0.4829 & 3.9$\times$10$^{1}$ &   19 &      49$\pm$12  &     0.84 $\pm$ 0.02	  & 8.1	  & 661 & 7.0	& $6^{+1}_{-1}$ & 4.72$-$7.22 & 3.5 \\
0529$+$075 &        0.0822 & 1.7$\times$10$^{3}$ &   58 &      11$\pm$2   &     0.93 $\pm$ 0.02		& 8.4	  & 460 & 3.3	& $26^{+7}_{-6}$ & 1.81$-$1.86 & 1.7\\
1045$-$188 &        0.0678 & 1.1$\times$10$^{3}$ &   48 &      17$\pm$3   &     0.94 $\pm$ 0.02	 	& 10.8	& 355 & 4.1	& $11^{+2}_{-2}$ & 2.44$-$2.71 & 2.1\\
1334$-$127 &        0.6867 & 1.2$\times$10$^{3}$ &   38 &      52$\pm$14  &     0.82 $\pm$ 0.01		& 7.5	  & 582 & 7.2	& $11^{+2}_{-2}$ & 4.68$-$7.43 & 3.6\\
1800$+$440 &        0.1871 & 9.9$\times$10$^{2}$ &   51 &      29$\pm$6   &     0.89 $\pm$ 0.01		& 6.6	  & 486 & 5.3	& $25^{+3}_{-3}$ & 3.00$-$3.10 &2.7\\
1849$+$670 &        0.4859 & 2.2$\times$10$^{3}$ &   18 &      18$\pm$4   &     0.84 $\pm$ 0.02		& 3.7	  & 2156 & 4.2	& $114^{+15}_{-17}$ & 2.21$-$2.23 & 2.2\\
2155$-$152 &        0.0255 & 1.6$\times$10$^{3}$ &   52 &      10$\pm$2   &     0.99 $\pm$ 0.03		& 6.1	  & 793 & 3.1	& $55^{+10}_{-13}$ & 1.69$-$1.72 & 1.7\\
2201$+$315 &        0.2438 & 1.5$\times$10$^{3}$ &   27 &      29$\pm$7   &     0.87 $\pm$ 0.03	 	& 9.9	  & 254 & 5.4	& $9^{+1}_{-2}$ & 3.47$-$5.67 & 2.7\\
2216$-$038 &        0.0428 & 2.9$\times$10$^{3}$ &   49 &       9$\pm$1   &     0.97 $\pm$ 0.02	 	& 15.8	& 445 & 3.0	& $7^{+1}_{-2}$ & 1.84$-$2.06 & 1.6\\
2345$-$167 &   $<$  0.1227 & 7.5$\times$10$^{2}$ &   44 & $<$  21         & $>$ 0.91	  		      & $>$ 7.5   & ... & 4.7	& 22 & ... & ...\\
\enddata

\tablecomments{\tiny{Columns are as follows:\\
(1) IAU name (B1950.0)\\
(2)	Ratio of the inverse Compton to synchrotron luminosities\\
(3) Volume of the synchrotron emission region in kpc$^3$\\
(4) Minimum energy magnetic field in $\mu$G for the case where there is no Doppler boosting, given by Eq. 1\\
(5) K is a function of observables and assumed quantities given by Eq. 3\\
(6) Radio to X-ray spectral index (1.4 GHz to 1 keV)\\
(7) Angle to line of sight with respect to the jet axis in ($^\circ$), assuming no deceleration or bending between the pc and kpc scales\\
(8) Deprojected jet length of the radio jet (core to hotspot) in kpc\\
(9) Doppler beaming parameter, assuming no deceleration or bending between the pc and kpc scales\\
(10) Jet bulk Lorentz factor assuming no deceleration or bending between pc and kpc scales\\
(11) Kpc-scale jet bulk Lorentz factor allowing for deceleration but no bending between pc and kpc scales\\
(12) Minimum kpc-scale jet bulk Lorentz factor, allowing for deceleration and bending between pc and kpc scales}}
\label{table:bmp}
\end{deluxetable}


\clearpage

\begin{thebibliography}{}

\bibitem[Angel \& Stockman(1980)]{AS80} {{Angel}, J.~R.~P. \& {Stockman}, H.~S.} 1980, {\araa}, 18, 321 

\bibitem[Blackburn(1995)]{BB95} Blackburn, J.\ K.\ 1995 in ASP Conf. Ser., Vol. 77 Astronomical Data Analysis Software and Systems IV, ed. R.\ A.\ Shaw, H.\ E.\ Payne, and J.\ J.\ E.\ Hayes (San Fransisco ASP), 367

\bibitem[Cara \& Lister(2008)]{CL08} {{Cara}, M. and {Lister}, M.~L.}, 2008, {\apj}, 674, 111

\bibitem[Cooper et al.(2007)]{NC07}{{Cooper}, N.~J., {Lister}, M.~L., {Kochanczyk}, M.~D.} 2007, \apjs, 171, 376

\bibitem[Cooper et al.(2009)]{NC09} {{Cooper}, N.~J., {Lister}, M.~L., {Kochanczyk}, M.~D.}, 2009, {VizieR Online Data Catalog}, 217, {10376-+}
 
\bibitem[Cooper(2010)]{NC10} Cooper, N., 2010, Ph.D. Thesis, Purdue University 
 
\bibitem[Conway \& Murphy(1993)]{CM93} {{Conway}, J.~E. \& {Murphy}, D.~W.} 1993, {\apj}, 411, 89

\bibitem[Coppi et al.(2002)]{CO02} Coppi et al. AAS Head Meeting \#5, \#26.19; Bulletin of the AAS, Vol 32, p.1226

\bibitem[Dermer \& Atoyan(2002)]{DA02} Dermer, C., D. \& Atoyan, A. M. 2002, \apj, 586, L81
 
\bibitem[Fanaroff \& Riley(1974)]{FR74} Fanaroff, B.\ L. \& Riley, J.\ M.\ 1974, \mnras, 167, 31

\bibitem[Fossati(1999)]{FO99} Fossati, G., Celotti, A., Ghisellini, G., Maraschi, L. 1999, Astronomical Society of the Pacific Conference Series, 159, 351

\bibitem[Ghisellini \& Tavecchio(2008)]{GT08} {{Ghisellini}, G. \& {Tavecchio}, F.} 2008, \mnras, 387, 1669

\bibitem[Georganopoulos \& Kazanas(2004)]{GK04} {{Georganopoulos}, M. \& {Kazanas}, D.} 2004, {\apjl}, 604, {L81-L84}

\bibitem[Harris \& Krawczynski(2002)]{HK02} Harris, D.\ E.\ \& Krawczynski, H.\ 2002, \apj, 565, 244

\bibitem[Harris \& Krawczynski(2006)]{HK06} Harris, D.\ E.\ \& Krawczynski, H.\ 2006, {\araa}, 44, 463

\bibitem[Hardcastle et al.(2002)]{HC02} Hardcastle, M.\ J., Birkinshaw, M., Cameron, R.\ A., Harris, D.\ E., Lonney, L.\ W., Worrall, D.\ M. 2002, \apj, 581, 948

\bibitem[Hovatta et al.(2009)]{TH09} {{Hovatta}, T., {Valtaoja}, E., {Tornikoski}, M., {L{\"a}hteenm{\"a}ki}, A.} 2009, \aap, 494, 527

\bibitem[Jester et al.(2006)]{SJ06} Jester, S., Harris, D.\ E., Marshall, H.\ L., Meisenheimer, K.\ 2006, \apj, 648, 900

\bibitem[Jorstad et al.(2004)]{SJ04} Jorstad, S.\ G., Marscher, A.\ P., Lister, M.\ L., Stirling, A.\ M., Cawthorne, T.\ V., G\'omez, J.\ L., Gear, W.\ K. 2004 \aj, 127, 3115

\bibitem[Jorstad et al.(2005)]{SJ05}{{Jorstad}, S.~G., {Marscher}, A.~P., {Lister}, M.~L., 
	{Stirling}, A.~M., {Cawthorne}, T.~V., {Gear}, W.~K., 
	{G{\'o}mez}, J.~L., {Stevens}, J.~A., {Smith}, P.~S., 
	{Forster}, J.~R., {Robson}, E.~I.} 2005, \aj, 130, 1418

\bibitem[Jorstad \& Marscher(2006)]{JM06} Jorstad, S.\ G. \& Marscher, A.\ P., 2006, Astronomische Nachrichten, 327, 227

\bibitem[Kharb et al.(2008)]{PK08} {{Kharb}, P., {O'Dea}, C.~P., {Baum}, S.~A., {Daly}, R.~A., {Mory}, M.~P., {Donahue}, M., {Guerra}, E.~J.}, 2008, {\apjs}, 174, 74

\bibitem[Kharb et al.(2010)]{PK10} Kharb, P., Lister, M. L., Cooper, N. J., 2010, ApJ, 710, 764

\bibitem[Linfield \& Perley(1984)]{LP84} Linfield, R. \& Perley, R.\ 1984, \apj, 279, 60

\bibitem[Lister et al.(2009a)]{ML09a} Lister M.\ L., Aller, H.\ D., Aller, M.\ F.,  Cohen, M.\ H., Homan, D.\ C., Kadler, M., Kellermann, K.\ I., Kovalev, Y.\ Y., Ros, E., Savolainen, T., Zensus, J.\ A., Vermeulen, R.\ C. 2009, \aj, 137, 3718

\bibitem[Lister et al.(2009b)]{ML09b}{{Lister}, M.~L., {Cohen}, M.~H., {Homan}, D.~C., {Kadler}, M., 
	{Kellermann}, K.~I., {Kovalev}, Y.~Y., {Ros}, E., {Savolainen}, T., 
	{Zensus}, J.~A.} 2009, \aj, 138, 1874

\bibitem[Lister \& Marscher(1997)]{LM97}{{Lister}, M.~L. \& {Marscher}, A.~P.} 1997, {\apj}, 476, {572}

\bibitem[Marshall et al.(2005)]{HM05} Marshall, H.\ L., Schwartz, D.\ A., Lovell, J.\ E.\ J., Murphy, D.\ W., Worrall, D.\ M. Birkinshaw, M., Gelbord, J.\ M., Perlman, E.\ S., Jauncey, D.\ L.\ 2005, \apjs, 156, 13

\bibitem[Moore et al.(1981)]{PM81} {{Moore}, P.~K., {Browne}, I.~W.~A., {Daintree}, E.~J., {Noble}, R.~G., {Walsh}, D.} 1981, {\mnras}, 197, 325

\bibitem[Mullin \& Hardcastle(2009)]{LM09} {{Mullin}, L.~M. and {Hardcastle}, M.~J.}, 2009, {\mnras}, 398, 1989

\bibitem[Pacholczyk(1970)]{AP70} {Pacholczyk}, A.~G. 1970, {Radio astrophysics. Nonthermal processes in galactic and extragalactic sources},

\bibitem[Padovani \& Urry(1992)]{PU92} {{Padovani}, P. \& {Urry}, C.~M.}, {\apj}, 387, 449

\bibitem[Sambruna et al.(2004)]{RS04} Sambruna, R.\ M., Gambill, J.\ K., Maraschi, L., Tavecchio, F., Cerutti, R., Cheung, C.\ C., Urry, C.\ M., Chartas, G.\ 2004, \apj, 608, 698

\bibitem[Tavecchio et al.(2001)]{TV01} {{Tavecchio}, F.,{Maraschi}, L., {Sambruna}, R.\ M., {Urry}, C.\ M.} 2001, Astronomical Society of the Pacific Conference Series, 234, 465

\bibitem[Tavecchio (2007)]{TV07} {{Tavecchio}, F.} 2007, {\apss}, 311, 247

\bibitem[Urry \& Padovani(1995)]{UP95} Urry, C.\ M.\ \& Padovani, P.\ 1995, \pasp, 107, 803 

\bibitem[WEBT(2007)]{WE07}{{WEBT Collaboration: W.~Collmar}, {B{\"o}ttcher}, M., 
	{Krichbaum}, T., {Bottacini}, E., {Burwitz}, V., {Cucchiara}, A., 
	{Grupe}, D., {Gurwell}, M., {Kretschmar}, P., {Pottschmidt}, K., 
	{Bremer}, M., {Leon}, S., {Ungerechts}, H., {Giommi}, P., 
	{Capalbi}, M.}, 2007, {ArXiv e-prints}, {0710.1096}
	
\bibitem[Wilson et al.(2001)]{WA01} Wilson, A.\ S., Young A.\ J., Shopbell, P.\ L.\ 2001, \itshape{Particles and Fields in Radio Galaxies Conference}\upshape, 250, 213

\bibitem[Worrall(2009)]{WD09} Worrall, D.\ M. 2009, \aapr, 17, 1

\end{thebibliography}
\end{document}